\documentclass[submission,copyright,creativecommons]{eptcs}
\providecommand{\event}{QPL 2025}

\usepackage{amsfonts}
\usepackage{amssymb}
\usepackage{amsthm}
\usepackage{iftex}
\usepackage{mathtools}
\usepackage{quiver}
\usepackage{subcaption}
\usepackage{xcolor}

\usepackage{hyperref}
\usepackage[capitalise]{cleveref}
\crefname{algorithm}{Alg.}{Algs.}
\crefname{section}{Sec.}{Secs.}
\crefname{definition}{Def.}{Defs.}
\crefname{table}{Tab.}{Tabs.}
\crefname{example}{Ex.}{Exs.}
\crefname{proposition}{Prop.}{Props.}
\crefname{theorem}{Thm.}{Thms.}
\crefname{corollary}{Cor.}{Cors.}
\crefname{appendix}{Appx.}{Appxs.}

\theoremstyle{plain}
\newtheorem{theorem}{Theorem}[section]
\newtheorem{lemma}[theorem]{Lemma}
\newtheorem{proposition}[theorem]{Proposition}
\newtheorem*{proposition*}{Proposition}
\newtheorem{corollary}[theorem]{Corollary}
\theoremstyle{definition}
\newtheorem{definition}[theorem]{Definition}

\newtheorem{example}[theorem]{Example}
\theoremstyle{remark}

\ifpdf
  \usepackage{underscore}
  \usepackage[T1]{fontenc}
\else
  \usepackage{breakurl}
\fi

\newcommand{\interp}[1]{[\![#1]\!]}

\title{Enriched Categories for Parameterized Circuit Semantics}

\author{Scott Wesley
    \institute{Dalhousie University\\ Halifax, Canada}}

\def\titlerunning{Enriched Categories for Parameterized Circuit Semantics}
\def\authorrunning{S. Wesley}

\begin{document}
\maketitle

\begin{abstract}
    It is well-known that combinatorial circuits are modeled mathematically by string diagrams in a monoidal category.
    Given a gate set $\Sigma$, the circuits over $\Sigma$ can be thought of as string diagrams in the free monoidal category generated by $\Sigma$.
    In this model, circuit semantics are then given by monoidal functors out of this free category.
    For quantum circuits, this functor is often valued in the category of unitary matrices.
    This model suffices for concrete quantum circuits, but fails to describe parameterized families of quantum circuits, such as those which arise in the analysis of ansatz circuits.
    Intuitively, this functor should be valued in parameterized families of unitary matices, though it is not immediately clear what this mean through a categorical lens.
    In this paper, we show that the parameterized semantics studied in prior work can be understood through enrichment and internal constructions.
    We determine sufficient conditions under which this construction yields a symmetric monoidal category, and suggest how these semantics could be extended to classical circuit analysis and parameterized equivalence checking.
\end{abstract}

\section{Introduction}

Combinatorial circuits map inputs to outputs without retaining any internal state.
These circuits have many important application, such as in complexity theory~\cite{Vollmer1982}, hardware design~\cite{HarrisWeste2010}, and quantum program compilation~\cite{CrossJavadiAbhari2022}.
This wide range of applications has motivated the mathematical analysis of combinatorial circuits.
One early approach to this problem was to view combinatorial circuits as string diagrams in a monoidal category~\cite{Hotz1965}.
This perspective has had many successes, such as providing decision procedures for the equality of certain classes of classical circuits~\cite{Lafont2003}, and optimization procedures for certain classes of quantum circuits~(see~\cite{Backens2021} for an introduction).

With the advent of quantum machine learning, new challenges have emerged in the field of combinatorial circuit analysis.
As in classical machine learning, a quantum machine learning model is a tensor network (i.e., circuit) with one or more free parameters~(see~\cite{HugginsPatil2019}).
The challenge in modeling these circuits categorically comes from the parameterized semantics.
In categorical circuit analysis, the choice of a semantic model is given by a monoidal functor from the category of circuits to the category of semantic interpretations.
For example, quantum circuit semantics are often described by monoidal functors valued in the category of finite-dimensional Hilbert spaces.
Several papers have attempted to give semantics to these parameterized circuits.
In~\cite{JeandelPerdrix2018}, the semantic functor was valued in continuous functions from $\mathbb{R}$ to Hilbert spaces of unitary matrices, though the monoidal structure on these continuous functions was never discussed explicitly.
In contrast,~\cite{MillerBakewell2020} circumvents the problem of semantics entirely, by valuing the semantic functor in the category of matrices over complex Laurent polynomials.
In both of these solutions, the parameterization is hidden from the categorical structure, and the parameter-free semantics are restricted to matrices.
As a result, it is unclear when these categorical methods are compatible with non-categorical results, such as the analytic theorems derived in~\cite{PehamBurgholzer2022}.

In this paper, we give a categorical account of parameterized semantics through the perspective of enriched category theory.
Assume that $\mathcal{C}$ is the category of semantic interprations, and $P$ is the collection of parameters.
Then evidently, $P$ must be an object in some other category $\mathcal{V}$.
Then parameterized maps from $X$ to $Y$ should be morphisms for the form $P \to \mathcal{C}( X, Y )$.
This only makes sense if $\mathcal{C}$ is enriched over $\mathcal{V}$.
We show that if $\mathcal{V}$ is Cartesian, then these parameterized morphisms form a category.
Moreover, if $\mathcal{C}$ is (symmetric) monoidal, then this parameterized category is as well.
This gives a general construction for parameterized circuit semantics, which agrees with the parameterized circuits in prior work.

The paper proceeds as follows.
In~\cref{Sect:Background}, we give a review of enirchment and monoidal categories from the perspective of string diagrams.
In~\cref{Sect:Construction}, we give a general construction for parameterized semantics.
In~\cref{Sect:Embed}, we show how the parameter-free interpretations embed into the parameterized semantics.
Many of the proofs in this paper are diagrammatic and have been relegated to appendices due to lack of space.

\section{Diagrams, Monoidal Categories, and Enrichment}
\label{Sect:Background}

We assume familiarity with monoidal categories and their string diagrams.
In particular, we assume that the reader is familiar with the characterization of a Cartesian monoidal category as symmetric monoidal category with a terminal unit $\mathbb{I}$, a uniform copying transformation $\Delta: ( - ) \Rightarrow ( - ) \otimes ( - )$, and a compatible uniform deleting transformation $e: ( - ) \Rightarrow \mathbb{I}$.
We also assume that the reader is familiar with meet semillatices as a family of Cartesian monoidal categories.
All preliminary material can be found in~\cref{Append:Cat} and~\cref{Append:MonCat}.
The rest of this section recalls enriched categories and enriched monoidal categories from the perspective of string diagrams.

\subsection{Enriched Category Theory}

In locally small categories, morphism composition is defined by functions between sets.
However, homsets often have additional structure, and composition often respects this structure.
For example, linear transformations form a vector space, with their composition given by a bilinear transformation.
This additional structure is studied through enriched category theory.
As shown later in this paper, enriched category theory also provides a framework to study parameterized families of morphisms.

\begin{definition}[{\cite{Kelly1982}}]
    Let $( \mathcal{V}, \otimes, \mathbb{I}, \alpha, \lambda, \rho )$ be a monoidal category.
    A $\mathcal{V}$-category $\mathcal{C}^{\mathcal{V}}$ consists of the following data.
    \begin{enumerate}
    \item \textbf{Objects}.
          A collection $\mathcal{C}^{\mathcal{V}}_0$.
    \item \textbf{Morphisms}.
          For each $X, Y \in \mathcal{C}^{\mathcal{V}}_0$, an object $\mathcal{C}^{\mathcal{V}}( X, Y ) \in \mathcal{V}_0$.
    \item \textbf{Identities}.
          For each $X \in \mathcal{C}^{\mathcal{V}}_0$, a generalized element $1^{\mathcal{V}}_X: \mathbb{I} \to \mathcal{C}^{\mathcal{V}}( X, X )$.
    \item \textbf{Composition}.
          For each $X, Y, Z \in \mathcal{C}^{\mathcal{V}}_0$, a morphism $M_{\mathcal{C}}: \mathcal{C}^{\mathcal{V}}( Y, Z ) \otimes \mathcal{C}^{\mathcal{V}}( X, Y ) \to \mathcal{C}^{\mathcal{V}}( X, Z )$.
    \end{enumerate}
    This data is subject to the condition that the following equations hold for all $X, Y, Z, W \in \mathcal{C}_0$.
    \begin{center}
        \input{figs/v_assoc_1}
\qquad
=
\qquad
\input{figs/v_assoc_2}

        \\
        \input{figs/v_unitality_1}
\;
=
\qquad
\input{figs/v_unitality_2}
\qquad
=
\qquad
\input{figs/v_unitality_3}

    \end{center}
\end{definition}

\begin{definition}[{\cite{Kelly1982}}]
    Let $( \mathcal{V}, \otimes, \mathbb{I}, \alpha, \lambda, \rho )$ be a monoidal category.
    A \emph{$\mathcal{V}$-functor} $F^{\mathcal{V}}: \mathcal{C}^{\mathcal{V}} \to \mathcal{D}^{\mathcal{V}}$ is a choice of objects $F_0( X ) \in \mathcal{D}^{\mathcal{V}}_0$ for each $X \in \mathcal{C}^{\mathcal{V}}_0$ and an indexed family $F_{X,Y}: \mathcal{C}( X, Y ) \to \mathcal{D}( F_0( X ), F_0( Y ) )$ of morphisms in $\mathcal{V}$ such that the following equations hold for each $X, Y, Z \in \mathcal{C}_0$.
    \begin{center}
        \input{figs/v_fcomp_1}
\qquad
=
\qquad
\input{figs/v_fcomp_2}

        \qquad\qquad\qquad
        \input{figs/v_funitality_1}
\quad\qquad
=
\quad\qquad
\input{figs/v_funitality_2}

    \end{center}
\end{definition}

\begin{definition}[{\cite{Kelly1982}}]
    Let $( \mathcal{V}, \otimes, \mathbb{I}, \alpha, \lambda, \rho )$ be a monoidal category with $F^{\mathcal{V}}, G^{\mathcal{V}}: \mathcal{C}^{\mathcal{V}} \to \mathcal{D}^{\mathcal{C}}$.
    A $\mathcal{V}$-natural transformation $\eta^{\mathcal{V}}: F^{\mathcal{V}} \Rightarrow G^{\mathcal{V}}$ is an indexed family of morphisms $\eta_X: \mathbb{I} \to \mathcal{D}( F_0( X ), G_0( X ) )$ such that the following equation holds for all $X, Y \in \mathcal{C}_0$.
    \begin{center}
        \input{figs/v_nat_1}
        \qquad
        =
        \qquad
        \input{figs/v_nat_2}
    \end{center}
\end{definition}

\begin{example}
    \label{Ex:Enriched}
    The simplest example of an enriched category is a \textbf{Set}-enriched category, since the \textbf{Set}-enriched categories are precisely the locally small categories.
    A more interesting set of examples are the \textbf{Top}-enriched categories, whose morphisms form topological spaces, and whose composition functions are continuous maps between these spaces.
    A simple example of a \textbf{Top}-enriched category is \textbf{FVect}.
    This is because: (1) linear transformations form finite-dimensional vector spaces; (2) all finite-dimensional vector spaces are topological spaces; and (3) all linear transformations between finite-dimensional vector spaces are continuous.

\end{example}

It is easy to show that $\mathcal{V}$-categories and $\mathcal{V}$-functors form a category, denoted $\textbf{Cat}( \mathcal{V} )$.
One would hope that each $\mathcal{V}$-category $\mathcal{C}^{\mathcal{V}} \in \textbf{Cat}( \mathcal{V} )_0$ corresponds to some category $\mathcal{C} \in \textbf{Cat}$.
Indeed, each $\mathcal{V}$-category $\mathcal{C}^{\mathcal{V}}$ has an \emph{underlying category} $\mathcal{C}$ such that $\mathcal{C}_0 = \mathcal{C}^{\mathcal{V}}_0$ and $\mathcal{C}( X, Y ) = \mathcal{V}( \mathbb{I}, \mathcal{C}^{\mathcal{V}}( X, Y ) )$.
In other words, the morphisms in $\mathcal{C}( X, Y )$ are the generalized elements of $\mathcal{C}^{\mathcal{V}}( X, Y )$ in $\mathcal{C}$.
Given $f \in \mathcal{C}( X, Y )$ and $g \in \mathcal{C}( Y, Z )$, their composite $g \circ f \in \mathcal{C}( X, Z )$ is defined to be $M_{\mathcal{C}} \circ ( g \otimes f ) \circ \rho_{\mathbb{I}}$.
It follows that each $\mathcal{V}$-functor $F^{\mathcal{V}}: \mathcal{C}^{\mathcal{V}} \to \mathcal{D}^{\mathcal{V}}$ induces a functor $F: \mathcal{C} \to \mathcal{D}$, and each $\mathcal{V}$-natural transformation $\eta^{\mathcal{V}}: F^{\mathcal{V}} \Rightarrow G^{\mathcal{V}}$ induces a natural transformation $\eta: F \Rightarrow G$.
This yields a functor $U: \textbf{Cat}( \mathcal{V} ) \to \textbf{Cat}$, as described in~\cite{Kelly1982} (for succinctness, we simply omit the superscript).

When $\mathcal{V}$ is a symmetric monoidal, the monoidal product for $\mathcal{V}$ extends to a monoidal product for $\textbf{Cat}( \mathcal{V} )$.
We follow the construction in~\cite{Kelly1982}.
Let $\mathcal{C}^{\mathcal{V}} \in \textbf{Cat}( \mathcal{V} )$ and $\mathcal{D}^{\mathcal{V}} \in \textbf{Cat}( \mathcal{V} )$.
The objects in $\mathcal{C}^{\mathcal{V}} \times \mathcal{D}^{\mathcal{V}}$ are taken to be $( \mathcal{C} \times \mathcal{D} )_0$.
Concretely, this means that for each $X \in \mathcal{C}^{\mathcal{V}}_0$ and $Y \in \mathcal{D}^{\mathcal{V}}_0$, the tuple $( X, Y )$ is an element in $( \mathcal{C} \times \mathcal{D} )_0$.
Then for each $X, Y \in \mathcal{C}^{\mathcal{V}}_0$ and $X', Y' \in \mathcal{D}^{\mathcal{V}}_0$, the morphisms in $( \mathcal{C}^{\mathcal{V}} \times \mathcal{D}^{\mathcal{V}} )( ( X, X' ), ( Y, Y' ) )$ are defined to be $\mathcal{C}^{\mathcal{V}}( X, Y ) \otimes \mathcal{D}^{\mathcal{V}}( X', Y' )$.
Then for each $X \in \mathcal{C}^{\mathcal{V}}_0$ and $X' \in \mathcal{D}^{\mathcal{V}}_0$, the identity for $( X, Y )$ is $( 1_X^{\mathcal{V}} \otimes 1_Y^{\mathcal{V}} ) \circ \lambda_{\mathbb{I}}$.
Composition in $\mathcal{C}^{\mathcal{V}} \times \mathcal{D}^{\mathcal{V}}$ then reduces to composition in $\mathcal{C}^{\mathcal{V}}$, $\mathcal{D}^{\mathcal{V}}$ via the symmetric braiding for $\mathcal{V}$, as illustrated by the following diagram.
\begin{center}
    \input{figs/vcat_prod}
\end{center}

\subsection{Enriched Monoidal Categories}

In this section, $\mathcal{V}$ is assumed to be a symmetric monoidal category.
As outlined in the previous section, this means that $\textbf{Cat}( \mathcal{V} )$ admits a monoidal product.
Given this product, the definition of a monoidal category extends readily to the enriched setting.
Intuitively, a $\mathcal{V}$-monoidal category $\mathcal{C}^{\mathcal{V}}$ is a monoidal structure on $\mathcal{C}$ which lifts to the enriched setting.
For example, the Kronecker tensor product on $\textbf{FVec}$ is bilinear, and consequently $\textbf{FVec}$ is a $\textbf{FVec}$-enriched category.

\begin{definition}[{\cite{KongZheng2018}}]
    Let $\mathcal{V}$ be a smc.
    A $\mathcal{V}$-monoidal category $( \mathcal{C}^{\mathcal{V}}, \otimes^{\mathcal{V}}, \mathbb{I}^{\mathcal{V}}, \alpha^{\mathcal{V}}, \lambda^{\mathcal{V}}, \rho^{\mathcal{V}} )$ is a $\mathcal{V}$-category $\mathcal{C}^{\mathcal{V}}$ with the following data.
    \begin{enumerate}
    \item \textbf{Monoidal Product}.
          A $\mathcal{V}$-functor $\mathcal{C}^\mathcal{V} \times \mathcal{C}^{\mathcal{V}} \to \mathcal{C}^{\mathcal{V}}$.
    \item \textbf{Monoidal Unit}.
          An object $\mathbb{I}^{\mathcal{V}} \in \mathcal{C}^{\mathcal{V}}_0$.
    \item \textbf{Associator}.
          A $\mathcal{V}$-natural isomorphism $\alpha^{\mathcal{V}}: ( - ) \otimes^{\mathcal{V}} ( ( - ) \otimes^{\mathcal{V}} ( - ) ) \Rightarrow ( ( - ) \otimes^{\mathcal{V}} ( - ) ) \otimes^{\mathcal{V}} ( - )$.
    \item \textbf{Unitors}.
          $\mathcal{V}$-natural isomorphism $\lambda^{\mathcal{V}}: ( \mathbb{I}^{\mathcal{V}} ) \otimes^{\mathcal{V}} ( - ) \Rightarrow ( - )$ and $\rho: ( - ) \otimes^{\mathcal{V}} ( \mathbb{I}^{\mathcal{V}} ) \Rightarrow ( - )$.
    \end{enumerate}
    This data is subject to the condition that $( \mathcal{C}, \otimes, \mathbb{I}, \alpha, \lambda, \rho )$ is a monoidal category.
\end{definition}

Since each $\mathcal{V}$-natural transformation unpacks to a family of equations in $\mathcal{V}$, then $\mathcal{V}$-natural transformations can be interpreted as equations between string diagrams in $\mathcal{V}$.
In particular, the $\mathcal{V}$-natural isomorphisms which define a $\mathcal{V}$-monoidal category also unpack to a family of diagrammatic equations.
A complete list of equations can be found in~\cref{Appendix:MonDiagrams}.

Whereas the definition of a $\mathcal{V}$-monoidal category was immediately obvious, the case of a braided $\mathcal{V}$-monoidal category requires more care.
This is because the braiding for $\mathcal{C}^{\mathcal{V}}$ involves both the underlying symmetry on \textbf{Cat} and induced symmetry on $\mathcal{V}$ as a symmetric monoidal category.
There exists several equivalent definitions in the literature of a braided $\mathcal{V}$-monoidal category, though the following definition is the most convenient for the constructions that follow.
The corresponding diagrams for this definition can be found in~\cref{Appendix:BrDiagrams}.

\begin{definition}[[\cite{KongZheng2018}]
    Let $\mathcal{V}$ be a smc with braiding $\sigma$.
    A \emph{commutative $\mathcal{V}$-functor} is a $\mathcal{V}$-functor $F^{\mathcal{V}}: \mathcal{C}^{\mathcal{V}} \times \mathcal{D}^{\mathcal{V}} \to \mathcal{E}^{\mathcal{V}}$ such that the data $F^{\mathrm{rev}}_0: ( x, y ) \mapsto F^{\mathcal{V}}_0( y, x )$ and $F^{\mathrm{rev}} = F \circ \sigma$ defines a $\mathcal{V}$-functor $F^{\textrm{rev}}: \mathcal{D}^{\mathcal{V}} \times \mathcal{C}^{\mathcal{V}} \to \mathcal{E}^{\mathcal{V}}$.
\end{definition}

\begin{definition}[\cite{KongZheng2018}]
    Let $\mathcal{V}$ be a smc and $( \mathcal{C}^{\mathcal{V}}, \otimes^{\mathcal{V}}, \mathbb{I}^{\mathcal{V}}, \alpha^{\mathcal{V}}, \lambda^{\mathcal{V}}, \rho^{\mathcal{V}} )$ be a monoidal $\mathcal{V}$-category such that $\otimes^{\mathcal{V}}$ commutative.
    A \emph{$\mathcal{V}$-braiding} for $\mathcal{C}^{\mathcal{V}}$ is a $\mathcal{V}$-natural isomorphism $\beta^{\mathcal{V}}: \otimes^{\mathcal{V}} \to \otimes^{\mathrm{rev}}$ such that $( \mathcal{C}, \otimes, \mathbb{I}, \alpha, \lambda, \rho, \beta )$ is a braided monoidal category.
    Moreover, if $( \mathcal{C}, \otimes, \mathbb{I}, \alpha, \lambda, \rho, \beta )$ is symmetric monoidal then $( \mathcal{C}^{\mathcal{V}}, \otimes^{\mathcal{V}}, \mathbb{I}^{\mathcal{V}}, \alpha^{\mathcal{V}}, \lambda^{\mathcal{V}}, \rho^{\mathcal{V}}, \beta^{\mathcal{V}} )$ is a \emph{symmetric $\mathcal{V}$-monoidal category}.
\end{definition}

\begin{example}
    \label{Ex:EnirchedMon}
    Recall from \cref{Ex:Enriched} that \textbf{FVect} is a \textbf{Top}-enriched category.
    Moreover, \textbf{FVect} is \textbf{Top}-monoidal with respect to the Kronecker tensor product.
    This is because the tensor product of matrices can be written as a multivariate polynomial function in the components of the two matrices, and is therefore continuous.
    Since permutations are continuous, \textbf{FVect} is also braided monoidal.
\end{example}

\section{Categories for Parameterized Semantics}
\label{Sect:Construction}

Fix a Cartesian category $( \mathcal{V}, \times, \mathbb{I}, \alpha, \rho, \lambda )$.
If $\mathcal{C}^{\mathcal{V}}$ is a $\mathcal{V}$-category, then the morphisms in its underlying category can be seen as generalized elements in $\mathcal{V}$.
Given some object $P \in \mathcal{V}_0$, one can then consider the generalized elements of $\mathcal{V}$ which factor through $P$, commonly known as the \emph{the $P$-shaped elements}.
In this section, $P$ will be thought of as the parameters used to construct parameterized families of morphisms in $\mathcal{C}^{\mathcal{V}}$.
First, it will be shown that using the Cartesian structure on $\mathcal{V}$, these $P$-shaped elements can be assembled into a category $\textbf{Param}( P, \mathcal{C} )$ which generalizes the parameterized constructions in~\cite{MillerBakewell2020,JeandelPerdrix2018}.
Second, it will be shown that if $\mathcal{C}^{\mathcal{V}}$ is $\mathcal{V}$-monoidal, then $\textbf{Param}( P, \mathcal{C} )$ is monoidal.
Third, it will be shown that if $\mathcal{C}^{\mathcal{V}}$ is braided (resp. symmetric), then so is $\textbf{Param}( P, \mathcal{C} )$.

\subsection{Constructing the Category}

\begin{figure}[t]
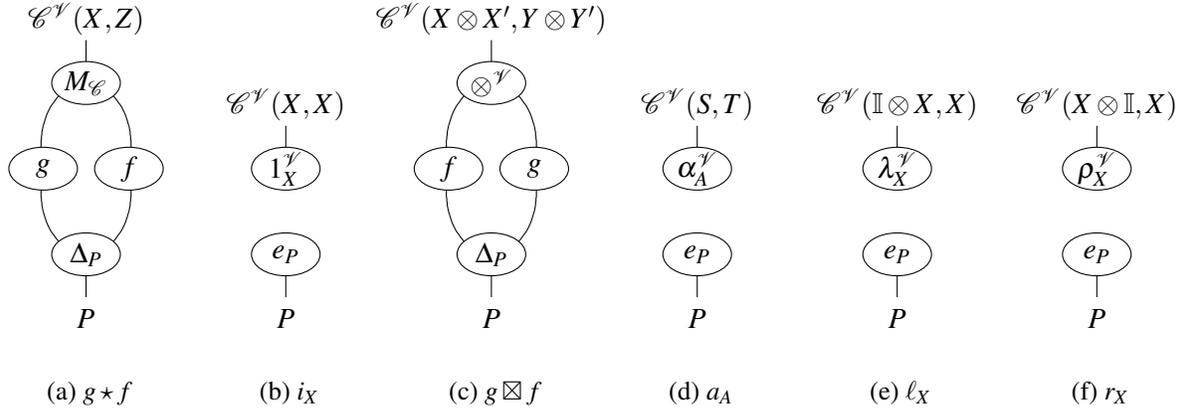

    \begin{subfigure}[b]{0.16\textwidth}
        \centering
        \input{figs/param_comp_2}
        \caption{$g \star f$}
    \end{subfigure}
    \begin{subfigure}[b]{0.16\textwidth}
        \centering
        \input{figs/param_id_2}
        \caption{$i_X$}
    \end{subfigure}
    \begin{subfigure}[b]{0.17\textwidth}
        \centering
        \input{figs/param_tensor_2}
        \caption{$g \boxtimes f$}
    \end{subfigure}
    \begin{subfigure}[b]{0.16\textwidth}
        \centering
        \input{figs/param_assoc_def}
        \caption{$a_A$}
    \end{subfigure}
    \begin{subfigure}[b]{0.16\textwidth}
        \centering
        \input{figs/param_lunit_def}
        \caption{$\ell_X$}
    \end{subfigure}
    \begin{subfigure}[b]{0.16\textwidth}
        \centering
        \input{figs/param_runit_def}
        \caption{$r_X$}
    \end{subfigure}
     \caption{The categorical and monoidal structure on $\mathbf{Param}( P, \mathcal{C} )$, depicted as string diagrams. In these diagrams $X,Y,Z \in \textbf{Param}( P, \mathcal{C} )$ with $f: X \to Y$, $g: Y \to Z$, and $h: X' \to Y'$. We write $Q = X \otimes ( Y \otimes Z )$, $R = ( X \otimes Y) \otimes Z$, and $A = ( X, Y, Z )$.}
    \label{Fig:Param1}
\end{figure}

This section provides the data for $\textbf{Param}( P, \mathcal{C} )$ and proves that this data satisfies the conditions of a category.
The objects in $\textbf{Param}( P, \mathcal{C} )$ are defined to be the objects in $\mathcal{C}$, since parameterized morphisms in $\textbf{Param}( P, \mathcal{C} )$ should have the same sources and targets as the morphisms in $\mathcal{C}$.
The morphisms from $X$ to $Y$ are defined to be the $P$-shaped elements of $\mathcal{C}^{\mathcal{V}}( X, Y )$, that is, elements of $\mathcal{V}( P, \mathcal{C}^{\mathcal{V}}( X, Y ) )$.
For each $X \in \mathcal{C}_0$, the unit $i_X$ for $X$ in $\textbf{Param}( P, \mathcal{C} )( X, X )$ is defined to be $1_X^{\mathcal{V}} \circ e_P$.
Given $f \in \textbf{Param}( P, \mathcal{C} )( X, Y )$ and $g \in \textbf{Param}( P, \mathcal{C} )( Y, Z )$, the composite $g \star f$ is defined to be $M_\mathcal{C} \circ ( g \times f ) \circ \Delta_P$, which first copies the parameter to pass to both of the morphisms, and then composes the morphisms in $\mathcal{C}^{\mathcal{V}}$.
Both $i_{(-)}$ and $( - ) \star ( - )$ are illustrated diagrammatically in~\cref{Fig:Param1}.

\begin{theorem}
    \label{Thm:ParamCat}
    $\mathbf{Param}( P, \mathcal{C} )$ is a category for each $P \in \mathcal{V}_0$.
    Moreover, $\mathbf{Param}( \mathbb{I}, \mathcal{C} ) = \mathcal{C}$.
\end{theorem}

\begin{proof}
    Let $P \in \mathcal{V}_0$.
    First, it must be shown that $( \star )$ is associative.
    Let $X \xrightarrow{f} Y \xrightarrow{g} Z \xrightarrow{h} W$ in $\mathbf{Param}( P, \mathcal{C} )$.
    Then the following equation of diagrams hold.
    \begin{center}
        $h \star ( g \star f )$
        \;
        =
        \;\;
        \input{figs/param_assoc_2}
        \;
        =
        \;\;
        \input{figs/param_assoc_3}
        \;
        =
        \;\;
        \input{figs/param_assoc_4}
        \;
        =
        \;
        $( h \star g ) \star f$
    \end{center}
    Since $f$, $g$, and $h$ were arbitrary, then $( \star )$ is associative.
    Next, it must be shown that $( \star )$ is unital.
    Let $f: X \to Y$ in $\mathbf{Param}( P, \mathcal{C} )$.
    Then the following equation of diagrams hold.
    \begin{center}
        $f \star i_x$
        \;\;
        =
        \;\;
        \input{figs/param_unit_1}
        \;\;
        =
        \;\;
        \input{figs/param_unit_2}
        \;\;
        =
        \;\;
        \input{figs/param_unit_3}
        \;\;
        =
        \;\;
        \input{figs/param_unit_4}
        \;\;
        =
        \;\;
        $f$
    \end{center}
    It follows by a symmetric argument that $i_y \star f = f$.
    Since $f$ was arbitrary, then $( \star )$ is unital.
    Since $P$ was arbitrary, then $\mathbf{Param}( P, \mathcal{C} )$ is a category for each $P \in \mathcal{V}_0$.
    Next, let $\mathcal{D} = \textbf{Param}( \mathbb{I}, \mathcal{C} )$.
    Clearly, $\mathcal{D}_0 = \mathcal{C}_0$ and $\mathcal{D}( X, Y ) = \mathcal{C}( X, Y )$ for each $X, Y \in \mathcal{C}_0$.
    Since $e_{\mathbb{I}} = 1_{\mathbb{I}}$, then $i_X = 1_X^{\mathcal{V}}$.
    Since $\Delta_{\mathbb{I}} = \rho_{\mathbb{I}}$, then $g \star f = M_{\mathcal{C}} \circ ( g \times f ) \circ \Delta_{\mathbb{I}} = M_{\mathcal{C}} \circ ( g \times f ) \circ \rho_{\mathbb{I}}$ for each $X \xrightarrow{f} Y \xrightarrow{g} Z$ in $\mathcal{C}$.
    In conclusion $\mathcal{D} = \mathcal{C}$.
\end{proof}

When $\mathcal{V}( \mathbb{I}, P ) \ne \varnothing$, each morphism $\theta \in \mathcal{V}( \mathcal{I}, P )$ can be thought of as a generalized parameter.
Then evaluating a parameterized morphism at $\theta$ is the same as composition with $\theta$.
In prior work, evaluation of parameterized functions could be proven to distribute over function composition~\cite{JeandelPerdrix2018,MillerBakewell2020}.
That is to say, if $M$ and $N$ are parameterized families of matrices, and $\theta$ is a parameter for the matrices, then $(M \circ N)( \theta ) = (M(\theta)) (N(\theta))$.
This equation suggests that evaluation should be functorial.
For each $\theta \in \mathcal{V}( \mathcal{I}, P )$, define the data for a functor $\textsf{ev}_\theta: \mathbf{Param}( P, \mathcal{C} ) \to \mathcal{C}$ to be $( \textsf{ev}_\theta )_0: X \mapsto X$ and $( \textsf{ev}_\theta )_{X,Y}: f \mapsto f \circ \theta$.
This is well-defined since $( \mathbf{Param}( P, \mathcal{C} ) )_0 = \mathcal{C}_0$ and $\mathcal{V}( \mathbb{I}, \mathcal{C}^{\mathcal{V}}( X, Y ) ) = \mathcal{C}( X, Y )$.
An important fact about $\textsf{ev}_{\theta}$ is that \emph{cancels out} composition with $e_P$.
It follows from this fact that $\textsf{ev}_{\theta}$ preserves many categorical structures, such as identities.

\begin{lemma}
    \label{Lem:InclEv}
    Let $\mathcal{V}$ be a Cartesian monoidal category with $P \in \mathcal{V}_0$ and $\mathcal{C}$ a $\mathcal{V}$-category.
    If $f \in \mathcal{C}( X, Y )$ and $\theta \in \mathcal{V}( \mathbb{I}, P )$, then $(\textsf{ev}_\theta)_{X,Y}( f \circ e_P ) = f$.
\end{lemma}

\begin{proof}
    Let $f \in \mathcal{C}( X, Y )$ and $\theta \in \mathcal{V}( \mathbb{I}, P )$.
    Then the following equation of diagrams holds.
    \begin{center}
        \;\;
        $(\textsf{ev}_\theta)_{X,Y}( f \circ e_P )$
        =
        \qquad
        \input{figs/ev_incl_1}
        \qquad
        =
        \qquad
        \input{figs/ev_incl_2}
        \qquad
        =
        \qquad
        \begin{tikzpicture}[baseline=(current bounding box.center),scale=0.56]
\useasboundingbox (-0.5,-0.5) rectangle (0.5,6.5);
\draw[,,] (0.00,0.00) -- (0.00,5.00);
\filldraw[fill=white] (0.00,0.00) ellipse (0.80cm and 0.50cm);
\draw (0.00,5.50) node{$\mathcal{C}^{\mathcal{V}}( X, Y )$};
\draw (0.00,0.00) node{$f$};
\end{tikzpicture}

        \qquad
        =
        \;\;
        $f$
    \end{center}
    Since $f$ and $\theta$ were arbitrary, then if $f \in \mathcal{C}( X, Y )$ and $\theta \in \mathcal{V}( \mathbb{I}, P )$.
\end{proof}

\begin{theorem}
    \label{Thm:CatEv}
    $\textsf{ev}_\theta: \mathbf{Param}( P, \mathcal{C} ) \to \mathcal{C}$ is a functor for each $\theta \in \mathcal{V}( \mathbb{I}, P )$.
\end{theorem}

\begin{proof}
    Fix some $\theta \in \mathcal{V}( \mathbb{I}, P )$.
    First, it must be show that $\textsf{ev}_\theta$ preserves morphism composition.
    Let $X \xrightarrow{f} Y \xrightarrow{g} Z$ in $\mathbf{Param}( P, \mathcal{C} )$.
    Then the following equation of diagrams hold.
    \begin{center}
        $( \textsf{ev}_\theta )_{X, Z}( g \star f )$
        \;\;
        =
        \quad\;
        \input{figs/ev_comp_1}
        \quad\;
        =
        \quad\;
        \input{figs/ev_comp_2}
        \quad\;
        =
        \;\;
        $( ( \textsf{ev}_\theta )_{Y,Z}( g ) ) \circ ( ( \textsf{ev}_\theta )_{X,Y}( f ) )$
    \end{center}
    Since $f$ and $g$ were arbitrary, then $\textsf{ev}_\theta$ preserves morphism composition.
    Next, it must be shown that $\textsf{ev}_\theta$ preserves identities.
    Let $X \in \mathcal{C}_0$.
    Since $i_X = 1_X \circ e_P$, then $( \textsf{ev}_\theta )_{X, X}( i_X ) = 1_X$ by \cref{Lem:InclEv}.
    Since  $X$ was arbitrary, then $\textsf{ev}_\theta$ preserves identities.
    Then $\textsf{ev}_\theta$ is a functor.
    Since $\theta$ was arbitrary, then $\textsf{ev}_\theta$ is a functor for each $\theta \in \mathcal{V}( \mathbb{I}, P )$.
\end{proof}

\begin{example}
    \label{Ex:ParamElt}
    Recall from \cref{Ex:EnirchedMon} that \textbf{FVect} is \textbf{Top}-enriched.
    Since \textbf{Top} is Cartesian, then there exists a category $\textbf{D} = \textbf{Param}( \mathbb{R}^2, \textbf{FVect} )$.
    This category provides semantics for quantum circuits in two variables.
    For any continuous map $f: \mathbb{R}^2 \to \mathbb{R}$, the $X$-rotation $R_X( f( \theta ) ) = \cos( f( \theta ) ) I + i \sin( f( \theta ) ) X$ is a morphism in $\mathcal{D}( 2, 2 )$.
    Moreover, $R_X( \theta_1 ) \star R_X( 2\theta_2 ) = R_X( \theta_1 + 2\theta_2 )$, as expected.
    Moreover, each $\theta \in \textbf{Top}( \mathbb{I}, \mathbb{R}^2 )$ picks out a parameter $( \theta_1, \theta_2 )$.
    If $\theta$ picks out $( \pi / 2, 0 )$, then $\textsf{ev}_\theta( R_X( \theta_1 ) \star R_X( 2\theta_2 ) ) = iX$.
\end{example}

\begin{example}
    \label{Ex:LatticeParam}
    In categories like \textbf{Set} and \textbf{Top}, objects can be thought of as sets and every non-empty set has generalized elements.
    However, there exists concrete Cartesian monoidal categories, such as meet semilattices which do not admit interesting generalized elements.
    Then in meet semilattices, it does not make sense to \emph{evaluate} prameterized morphisms.
    However, the construction still makes sense and yields interesting results.
    For example, consider the $3$-element poset $X = \{ \bot, 1, \top \}$ with the self-evident order $\bot < 1 < \top$.
    This poset forms a meet semilattice with $\land: X \times X \to X$ defined as follows.
    \begin{align*}
        \bot \land \bot &= \bot
        &
        \bot \land 1 &= \bot
        &
        \bot \land \top &= \bot
        \\
        1 \land \bot &= \bot
        &
        1 \land 1 &= 1
        &
        1 \land \top &= 1
        \\
        \top \land \bot &= \bot
        &
        \top \land 1 &= 1
        &
        \top \land \top &= \top
    \end{align*}
    Then $( X, \le, \land )$ defines a Cartesian monoidal category $\mathcal{V}$ where $\land$ is the monoidal product and $\top$ is the monoidal unit.
    Every $\mathcal{V}$-enriched category corresponds to a (possibly infinite) directed graph whose edges are labeled by elements of $X$, with pairs of composable edges $f: X \to Y$ and $g: Y \to Z$ satisfying $g \land f \le ( g \circ f)$.
    Moreover, each vertex $X$ must have a self-loop $X \to X$ labeled by the element $T$.
    This is because $1_X \in \mathcal{V}( \top, \mathcal{C}^{\mathcal{V}}( X, X )$ implies $\top \le \mathcal{C}^{\mathcal{V}}( X, X )$.
    As a simple example, consider a category $\mathcal{C}^{\mathcal{V}}$ such that $\mathcal{C}^{\mathcal{V}}_0 = \{ A, B, C, D \}$ and the homsets are defined as follows.
    \begin{align*}
        \mathcal{C}( A, A ) &= \top
        &
        \mathcal{C}( A, B ) &= \top
        &
        \mathcal{C}( A, C ) &= 1
        &
        \mathcal{C}( A, D ) &= \top
        \\
        \mathcal{C}( B, A ) &= \bot
        &
        \mathcal{C}( B, B ) &= \top
        &
        \mathcal{C}( B, C ) &= \bot
        &
        \mathcal{C}( B, D ) &= 1
        \\
        \mathcal{C}( C, A ) &= \bot
        &
        \mathcal{C}( C, B ) &= \bot
        &
        \mathcal{C}( C, C ) &= \top
        &
        \mathcal{C}( C, D ) &= \top
        \\
        \mathcal{C}( D, A ) &= \bot
        &
        \mathcal{C}( D, B ) &= \bot
        &
        \mathcal{C}( D, C ) &= \bot
        &
        \mathcal{C}( D, D ) &= \top
    \end{align*}
    The corresponding graph is illustrated below.
    
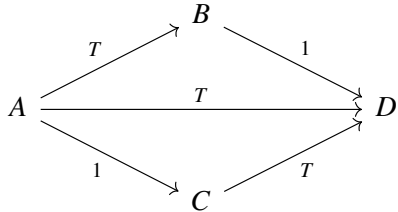
\begin{center}
\begin{tikzcd}
	&& B \\
	A &&&& D \\
	&& C
	\arrow["1", from=1-3, to=2-5]
	\arrow["T", from=2-1, to=1-3]
	\arrow["T", from=2-1, to=2-5]
	\arrow["1"', from=2-1, to=3-3]
	\arrow["T"', from=3-3, to=2-5]
\end{tikzcd}
    \end{center}
    Since $\mathcal{C}( X, Y ) = \mathcal{V}( \top, \mathcal{C}^{\mathcal{V}}( X, Y ) )$, then the underlying category $\mathcal{C}$ picks out the posetal category such that $X \rightarrow Y$ if and only if $\mathcal{C}( X, Y ) = \top$.
    The corresponding graph is illustrated below.
    \begin{center}
\begin{tikzcd}
	B &&& A &&& D &&& C
	\arrow[from=1-4, to=1-1]
	\arrow[from=1-4, to=1-7]
	\arrow[from=1-10, to=1-7]
\end{tikzcd}
    \end{center}
    From a logical perspective, $\mathcal{C}$ picks out the edges which as absolutely certain.
    Another reasonable construction is to first pick out a truth value $P \in X$, and then construct the posetal category containing the edges entailed by $P$.
    Since $( \textbf{Param}( P, \mathcal{C} ) )( X, Y ) = \mathcal{V}( P, \mathcal{C}( X, Y ) )$, then this is precisely the construction outlined in this section.
    For example, $\textbf{Param}( 1, \mathcal{C} )$ corresponds to the following graph.
    
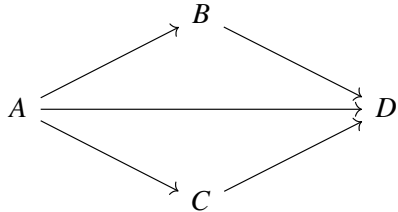
\begin{center}
\begin{tikzcd}
	&& B \\
	A &&&& D \\
	&& C
	\arrow["", from=1-3, to=2-5]
	\arrow["", from=2-1, to=1-3]
	\arrow["", from=2-1, to=2-5]
	\arrow["", from=2-1, to=3-3]
	\arrow["", from=3-3, to=2-5]
\end{tikzcd}
    \end{center}
    Moreover, $\textbf{Param}( \bot, \mathcal{C} )$ is the complete graph on $4$ vertices.
\end{example}

\subsection{Inheriting the Monoidal Structure}

Assume that $\mathcal{C}^{\mathcal{V}}$ admits a $\mathcal{V}$-monoidal structure $( \mathcal{C}^{\mathcal{V}}, \otimes^{\mathcal{V}}, \mathbb{J}^{\mathcal{V}}, \alpha^{\mathcal{V}}, \lambda^{\mathcal{V}}, \rho^{\mathcal{V}} )$.
This section shows how $\mathcal{D} = \mathbf{Param}( P, \mathcal{C} )$ inherits the monoidal structure of $\mathcal{C}^{\mathcal{V}}$.
The first step is to define a monoidal product $\boxtimes: \mathcal{D} \times \mathcal{D} \to \mathcal{D}$.
As expected, $\boxtimes_0 = \otimes_0$.
Then for each pair of functions $f: X \to Y$ and $g: X' \to Y'$ in $\mathcal{D}$, $f \boxtimes g$ is defined to be $\otimes^{\mathcal{V}}_{((X,X'),(Y,Y'))} \circ ( f \times g) \circ \Delta_P$, as illustrated in~\cref{Fig:Param1}.

\begin{lemma}
    \label{Lem:BoxFunc}
    $\boxtimes: \mathbf{Param}( P, \mathcal{C} ) \times \mathbf{Param}( P, \mathcal{C} ) \to \mathbf{Param}( P, \mathcal{C} )$ is a functor.
\end{lemma}

The next step is to show that $\boxtimes$ is unital and associative.
Since the objects of $\mathcal{D}$ share the same monoidal structure as $\mathcal{C}$, then the monoidal unit for $\boxtimes$ should be $\mathbb{J}$.
It remains to define the natural transformations $a: ( - ) \boxtimes ( ( - ) \boxtimes ( - ) ) \Rightarrow ( ( - ) \boxtimes ( - ) ) \boxtimes ( - )$, $\ell: \mathbb{J} \boxtimes ( - ) \to ( - )$ and $r: ( - ) \boxtimes \mathbb{J} \to ( - )$.
For each choice of $X, Y, Z \in \mathcal{C}_0$, define $a_{(X,Y,Z)}$ to be $\alpha^{\mathcal{V}}_{(X,Y,Z)} \circ e_P$.
Analogously, for each $X \in \mathcal{C}$, define $\ell_X$ to be $\lambda^{\mathcal{V}}_X \circ e_P$ and $r_X$ to be $\rho^{\mathcal{V}}_X \circ e_P$.
Of course, each component of $a$, $\ell$, and $r$ must be invertible, so that the transformations are isomorphisms.
Thankfully, composition with $e_P$ always preserves inverses.
Diagrammatic illustrations of $a$, $\ell$, and $r$ can be found in~\cref{Fig:Param1}.

\begin{lemma}
    \label{Lem:IncInv}
    If $X \xrightarrow{f} Y$ is invertible in $\mathcal{C}$, then $k = f \circ e_P$ is invertible in $\mathbf{Param}( P, \mathcal{C} )$.
\end{lemma}

\begin{proof}
    It suffices to show that $f^{-1} \circ e_P$ is the inverse to $k$.
    \begin{center}
        $( f^{-1} \circ e_p ) \star k$
        \;
        =
        \quad\;
        \input{figs/box_assoc_inv_1}
        \quad
        =
        \quad\;
        \input{figs/box_assoc_inv_2}
        \quad
        =
        \quad\;
        \input{figs/box_assoc_inv_3}
        \quad
        =
        \quad\;
        \input{figs/box_assoc_inv_4}
        \quad\;
        =
        \;
        $i_X$
    \end{center}
    Then $f^{-1} \circ e_P$ is the left inverse to $k$ in $\textbf{Param}( P, \mathcal{C} )$.
    By a symmetric argument, $f^{-1} \circ e_P$ is also the right inverse to $k$ in $\textbf{Param}( P, \mathcal{C} )$.
    In conclusion, $k$ is invertible in $\textbf{Param}( P, \mathcal{C} )$.
\end{proof}

\begin{lemma}
    \label{Lem:BoxAssoc}
    $a: ( - ) \boxtimes ( ( - ) \boxtimes ( - ) ) \Rightarrow ( ( - ) \boxtimes ( - ) ) \boxtimes ( - )$ is a natural isomorphism.
\end{lemma}

\begin{lemma}
    \label{Lem:BoxUnitors}
    $\ell: \mathbb{J} \boxtimes ( - ) \Rightarrow ( - )$ and $r: ( - ) \boxtimes \mathbb{J} \Rightarrow ( - )$ are natural isomorphisms.
\end{lemma}

It remains to be shown that $a$, $\ell$, and $r$ satisfy the coherence conditions of a monoidal category.
This is true, since $a$, $\ell$, and $r$ are defined to be the inclusions of $\alpha$, $\lambda$, and $\rho$ from $\mathcal{C}$ into $\textbf{Param}( P, \mathcal{C} )$.
This inclusion is explored more in~\cref{Sect:Embed}.
For now, this property is used to show that $\textbf{Param}( P, \mathcal{C} )$ also inherits the coherence equations from $\mathcal{C}$.

\begin{theorem}
    \label{Thm:ParamMon}
    $( \mathbf{Param}( P, \mathcal{C} ), \boxtimes, a, \ell, r )$ is a monoidal category for each $P \in \mathcal{V}_0$.
\end{theorem}

\begin{theorem}
    \label{Thm:ParamEv}
    $\textsf{ev}_\theta: \mathbf{Param}( P, \mathcal{C} ) \to \mathcal{C}$ is a strict monoidal functor for each $\theta \in \mathcal{V}( \mathbb{I}, P )$.
\end{theorem}

\begin{proof}
    Let $\theta \in \mathcal{V}( \mathbb{I}, P )$.
    Clearly $\textsf{ev}_\theta$ preserves the monoidal unit, since $(\textsf{ev}_\theta)_0$ is the identity.
    Next, it must be shown that $\textsf{ev}_\theta$ preserves monoidal composition.
    Let $X \xrightarrow{f} Y$ and $X' \xrightarrow{g} Y'$ in $\textbf{Param}( P, \mathcal{C} )$.
    Then the follow equation of diagrams holds.
    \begin{center}
        $( \textsf{ev}_\theta )_{X \boxtimes X', Y \boxtimes Y'}( g \boxtimes f )$
        =
        \quad
        \input{figs/ev_tensor_1}
        \qquad
        =
        \qquad
        \input{figs/ev_tensor_2}
        \quad
        =
        $( ( \textsf{ev}_\theta )_{X,Y}( g ) ) \otimes ( ( \textsf{ev}_\theta )_{X',Y'}( f ) )$
    \end{center}
    Since $f$ and $g$ were arbitrary, then $\textsf{ev}_\theta$ preserves monoidal composition.
    It remains to be shown that $\textsf{ev}_\theta$ preserves the structural natural transformations.
    Let $X, Y, Z \in \mathcal{C}_0$.
    Then it follows from \cref{Lem:InclEv} that $\textsf{ev}_\theta( a_{(X,Y,Z)} ) = \alpha_{(X,Y,Z)}$, $\textsf{ev}_\theta( \ell_X ) = \lambda_X$, and $\textsf{ev}_\theta( r_X ) = \rho_X$.
    Since $X$, $Y$, and $Z$ were arbitrary, then $\textsf{ev}_\theta$ preserves the structural natural transformations of $\mathcal{C}$.
    In conclusion, $\textsf{ev}_\theta$ is a strict monoidal functor.
\end{proof}

\begin{example}
    \label{Ex:ParamMon}
    Recall the example of $\mathcal{D} = \textbf{Param}( \mathbb{R}^2, \textbf{FVect} )$ from \cref{Ex:ParamElt}.
    Since \textbf{FVect} is \textbf{Top}-monoidal with respect to the Kronecker tensor product $\otimes$, then $\mathcal{D}$ inherits this monoidal structure of $( \textbf{FVect}, \otimes, \mathbb{C}, \alpha, \lambda, \rho )$.
    For example, $R_X( \theta_1 ) \boxtimes R_X( 2 \theta_2 )$ is an element of $\mathcal{D}( 4, 4 )$.
    This operator corresponds to a $2$-qubit circuit, which rotates the first qubit by $\theta_1$ degrees, and the second qubit by $\theta_2$ degrees.
    If $\theta \in \textbf{Top}( \mathbb{I}, \mathbb{R}^2 )$ picks out $( \pi / 2, 0 )$, then $\textsf{ev}_\theta( R_X( \theta_1 ) \boxtimes R_X( 2 \theta_2 ) ) = iX \otimes I$.
\end{example}

\subsection{Inheriting the Monoidal Braiding}

Assume that $( \mathcal{C}^{\mathcal{V}}, \otimes^{\mathcal{V}}, \mathbb{J}^{\mathcal{V}}, \alpha^{\mathcal{V}}, \lambda^{\mathcal{V}}, \rho^{\mathcal{V}} )$ admits a $\mathcal{V}$-braiding $\beta^{\mathcal{V}}$.
This sections show how $\mathcal{D}$ inherits the monoidal structure of $\mathcal{C}^{\mathcal{V}}$.
The first step is to define a natural isomorphism $b: ( - ) \boxtimes ( - ) \Rightarrow ( - ) \boxtimes ( - ) \circ c$ where $c$ is the symmetric braiding on \textbf{Cat} as a Cartesian category.
As in the cases of $a$, $\ell$, and $r$, the solution is to simply define each $b_{(X,Y)}$ to be the inclusion $\beta_{(X,Y)}^{\mathcal{V}} \circ e_P$ of $\beta_{(X,Y)}^{\mathcal{V}}$ into $\mathcal{D}( X, Y )$.

Unfortunately, the naturality of $\beta^{\mathcal{V}}$ does not immediately imply the naturality of $b$.
This is because $b$ is natural with respect to $c$, whereas $\beta^{\mathcal{V}}$ is natural with respect to $\sigma$, the braiding on $\mathcal{V}$.
Of course, $\beta$ is natural with respect to $c$, though this naturality only concerns the generalized elements of $\mathcal{C}$, rather than the $P$-shaped elements of $\mathcal{C}$.
To handle $P$-shaped elements of $\mathcal{C}$, the naturality of $\beta$ must be combined with the naturality of $\sigma$, to introduce a $c$ term.
To introduce a $\sigma$ term, it suffices to recall that $\Delta = \sigma \circ \Delta$.

\begin{lemma}
    \label{Lem:BoxSym}
    $b: ( - ) \boxtimes ( - ) \Rightarrow ( - ) \boxtimes ( - ) \circ c$ is a natural isomorphism.
\end{lemma}

It remains to be shown that $b$ satisfies the coherence conditions for a monoidal braiding.
However, this part of the proof is straight-forward, and follows from the same techniques used to prove the coherence of $a$, $\ell$, and $r$.
That is, the coherence of $\beta$ yields several diagrammatic equations, which immediately imply the coherence for $b$, once all of the $a$, $b$, and $i$ terms have been expanded.
This yields a braiding on $\textbf{Param}( P, \mathcal{C} )$, which is symmetric provided $\beta^{\mathcal{V}}$ is symmetric.

\begin{theorem}
    \label{Thm:ParamBraid}
    $( \mathbf{Param}( P, \mathcal{C} ), \boxtimes, a, \creflabelformat{}{}, r, b )$ is a braided monoidal category.
    Moreover, if $\beta$ is symmetric, then $b$ is also symmetric.
\end{theorem}

\begin{corollary}
    \label{Cor:BraidEv}
    $\textsf{ev}_\theta: \mathbf{Param}( P, \mathcal{C} ) \to \mathcal{C}$ is a strict braided monoidal functor for each $\theta \in \mathcal{V}( \mathbb{I}, P )$.
\end{corollary}

\begin{proof}
    Let $\theta \in \mathcal{V}( \mathbb{I}, P )$.
    By \cref{Thm:ParamEv} the functor $\textsf{ev}_\theta$ is strict monoidal and by \cref{Thm:ParamBraid} the monoidal category $\textbf{Param}( P, \mathcal{C} )$ admits a braiding $b$.
    Let $X, Y \in \mathcal{C}_0$.
    Since $b_{X,Y} = \beta_{X,Y} \circ e_P$, then $\textsf{ev}_\theta( b_{X,Y} ) = \beta_{X,Y}$ by \cref{Lem:InclEv}.
    Since $X$ and $Y$ were arbitrary, then $\textsf{ev}_\theta$ preserves the braiding on $\textbf{Param}( P, \mathcal{C} )$.
    Then $\textsf{ev}_\theta$ is a strict braided monoidal functor.
    Since $\theta$ was arbitrary, then $\textsf{ev}_\theta$ is a strict braided monoidal functor for each $\theta \in \mathcal{V}( \mathbb{I}, \mathcal{C} )$.
\end{proof}

\begin{example}
    \label{Ex:PAramBraid}
    Recall the example of $\mathcal{D} = \textbf{Param}( \mathbb{R}^2, \textbf{FVect} )$ from \cref{Ex:ParamMon}.
    Since \textbf{FVect} is braided \textbf{Top}-monoidal, then $\mathcal{D}$ also inherits a braiding from \textbf{FVect}.
    For example the braiding on $\mathbb{C} \oplus \mathbb{C}$ is $b_{(1,1)} = \beta_{(1,1)} \circ e_P$.
    Unpacking this definition in \textbf{Top},
    \begin{equation*}
        b_{(1,1)}: ( \theta_1, \theta_2 )
        \mapsto
        \begin{bmatrix}
            0 & 1 \\
            1 & 0
        \end{bmatrix}.
    \end{equation*}
    This construction behaves as expected.
    For example,
    \begin{equation*}
        b_{(1,1)}( \theta ) \star R_Z( \theta_1 ) \star b_{(1,1)}( \theta )
        =
        \begin{bmatrix}
            0 & 1 \\
            1 & 0
        \end{bmatrix}
        \star
        \begin{bmatrix}
            e^{i\theta} & 0 \\
            0 & e^{-i\theta}
        \end{bmatrix}
        \star
        \begin{bmatrix}
            0 & 1 \\
            1 & 0
        \end{bmatrix}
        =
        \begin{bmatrix}
            e^{-i\theta} & 0 \\
            0 & e^{i\theta}
        \end{bmatrix}
        =
        R_Z( -\theta_1 ).
    \end{equation*}
    Intuitively, $b_{(1,1)}$ is the inclusion of $\beta_{(1,1)}$ into $\mathcal{D}( 1, 1 )$.
    This is made precise in the next section.
\end{example}

\section{Extending from Parameter-Free Semantics}
\label{Sect:Embed}

As in previous sections, let $\mathcal{V}$ be a smc and $\mathcal{C}$ be a $\mathcal{V}$-category.
In previous work~\cite{MillerBakewell2020,JeandelPerdrix2018,PehamBurgholzer2022,XuLi2022}, parameterized circuits are thought of as a generalization of parameter-free circuits.
That is to say, every parameter-free circuit is trivially a parameterized circuit.
Formally, if $C$ is a parameter-free circuit with semantic interpretation $\interp{C}$, then its parameterized interpretation should be a constant function valued at $\interp{C}$.
Indeed, there exists an embedding $j$ of $\mathcal{C}$ into $\mathcal{D} = \textbf{Param}( P, \mathcal{C} )$ such that each $\interp{C}$ maps to a \emph{constant function} $j( \interp{C} )$ which always evaluates to \interp{C}.

Stated categorically, this means that for each $\theta \in \mathcal{V}( \mathbb{I}, \mathcal{C} )$, the functor $\textsf{ev}_\theta$ is a retraction of $\mathcal{D}$ onto $\mathcal{C}$ in \textbf{Cat}.
Moreover, the corresponding section $j: \mathcal{C} \to \mathcal{D}$ is the same for all choices of $\theta$.
In this sense, $j$ is the inclusion of $\mathcal{C}$ into $\mathcal{D}$ such that each morphism in $\mathcal{C}$ maps to a constant function in $\mathcal{D}$.
The functor $j: \mathcal{C} \to \textbf{Param}( P, \mathcal{C} )$ is defined such that $j_0: X \mapsto X$ and for each $f \in \mathcal{C}( X, Y )$, $j_{X,Y}( f ) = f \circ e_P$.

\begin{theorem}
    \label{Thm:SemIncl}
    $j: \mathcal{C} \to \textbf{Param}( P, \mathcal{C} )$ is functorial.
    If $\mathcal{C}^{\mathcal{V}}$ is $\mathcal{V}$-monoidal, then $j$ is strict monoidal.
    If $\mathcal{C}^{\mathcal{V}}$ is $\mathcal{V}$-braided, then $j$ is strict braided monoidal.
    If $\mathcal{V}( \mathbb{I}, P ) \ne \varnothing$, then $j$ is faithful.
\end{theorem}

\begin{corollary}
    \label{Cor:Retract}
    For each $\theta \in \mathcal{V}( \mathbb{I}, \mathcal{C} )$, $\textsf{ev}_\theta$ is a retraction with $\textsf{ev}_\theta \circ j = 1_{\mathcal{C}}$.
\end{corollary}

\begin{proof}
    Let $\theta \in \mathcal{V}( \mathbb{I}, \mathcal{C} )$.
    It must be shown that $F = \textsf{ev}_\theta \circ j$ is the identity functor on $\mathcal{C}$.
    Clearly, $F_0$ is the identity on objects, since $F_( X ) = (\textsf{ev}_\theta)_0( j_0( X ) ) = (\textsf{ev}_\theta)_0( X ) = X$ for each $X \in \mathcal{C}_0$.
    Next, let $f: X \to Y$ in $\mathcal{C}$.
    Since $j_{X,Y}( f ) = f \circ e_P$, then $F_{X,Y}( f ) = f$ by \cref{Lem:InclEv}.
    Since $f$ was arbitrary, then $F$ is also the identity on homsets.
    Then $\textsf{ev}_\theta \circ j = 1_{\mathcal{C}}$.
    Since $\theta$ was arbitrary, then for each $\theta \in \mathcal{V}( \mathbb{I}, \mathcal{C} )$, $\textsf{ev}_\theta$ is a retraction with $\textsf{ev}_\theta \circ j = 1_{\mathcal{C}}$.
\end{proof}

\begin{example}
    Recall the example of $\mathcal{D} = \textbf{Param}( \mathbb{R}^2, \textbf{FVect} )$ from \cref{Ex:ParamMon}.
    For each choice of $n$ and $m$ in $\mathbb{N}$, the morphism $b_{(n,m)}$ is the inclusion of the permutation matrix $\beta_{(n,m)}$ into $\mathcal{D}( n, m )$.
    That is, $b_{(n,m)} = j\left( \beta_{(n,m)} \right)$.
    Since $\mathcal{V}( \mathbb{I}, \mathbb{R}^2 )$ is non-empty, then $j$ is also faithful.
    Indeed, if $j( f ) = j( g )$, then $f = g$ because $( \textsf{ev}_\theta )_{X,Y}( j( f ) ) = ( \textsf{ev}_\theta )_{X,Y}( j( g ) )$ where $\theta = (0, 0)$.
\end{example}

Since every retraction defines an idempotent, it is natural to ask how these idempotents act on $\textbf{Param}( P, \mathcal{C} )$.
Pick $\theta \in \mathcal{V}( \mathbb{I}, P )$ and define $\textsf{const}_\theta = j \circ \textsf{ev}_\theta$.
Then given any parameterized morphism $f: X \to Y$ in $\mathcal{D}$, the functor $(\textsf{const}_\theta)_{X,Y}( f )$ picks out the constant function in $\mathcal{D}( X, Y )$ that agrees with $f$ when evaluated at $\theta$.
Formally, this means that $\textsf{ev}_\kappa( \textsf{const}_\theta( f ) ) = \textsf{ev}_\theta( f )$ for all $\kappa \in \mathcal{V}( \mathbb{I}, P )$.
This is illustrated by the following equation of diagrams.
\begin{center}
    $\textsf{ev}_\kappa( \textsf{const}_\theta( f ) )$
    =
    \qquad
    \input{figs/const_eval_1}
    \qquad
    =
    \qquad
    \input{figs/const_eval_2}
    \qquad
    =
    \qquad
    \input{figs/const_eval_3}
    \qquad
    =
    $\textsf{ev}_\theta( f )$
    \vspace{10pt}
\end{center}

\begin{example}
    Recall the example of $\mathcal{D} = \textbf{Param}( \mathbb{R}^2, \textbf{FVect} )$ from \cref{Ex:ParamMon}.
    As before, let $\theta = ( \pi / 2, 0 )$.
    Since $\textsf{ev}_\theta( R_X( \theta_1 ) ) = iX$, then $\textsf{const}_\theta( R_X( \theta_1 ) ): ( \theta_1, \theta_2 ) \mapsto iX$.
\end{example}

\section{Discussion and Future Work}

This paper introduced a notion of parameterized morphism through enrichment, which characterizes the semantics of parameterized quantum circuits, and also more abstract logical constructions. 
To the best of our knowledge, this is the first work to consider categorical semantics for parameterized circuits.
There is ongoing work outlined in~\cite{KoziellPipe2024} to model parameterized circuits though lenses, though this work has not been published yet.
We propose several directions for future work.

\paragraph{Parameterized Equivalence Checking.}
An important problem in parameterized circuit analysis is parameterized equivalence checking.
Given two circuits $C_1$ and $C_2$ parameterized by $P$, this question asks if $\interp{C_1}( \theta ) = \interp{C_2}( \theta )$ for all $\theta \in P$.
An obvious direction for future work is to reconstruct the problem of parameterized equivalence in this more general framework.
It can then be asked which domain-specific equivalence checking techniques, such as the analytic methods of~\cite{PehamBurgholzer2022} generalize to wider classes of parameterized semantics.
It would also be interesting to rephrase standard techniques, such as reparameterization (see,~e.g.,~\cite{WateringYeung2024}), in this framework, to see if any new insights can be gained.

\paragraph{From Logic to Parameters.}
As outlined in~\cref{Ex:Heyting}, all Heyting semilattices are Cartesian categories.
This means that any categories enriched over a Heyting semilattices can be parameterized by valuations in the underlying model intuitionistic logic.
It would be interesting to expand~\cref{Ex:LatticeParam} in this direction, and to explore applications to program analysis.

\paragraph{Change of Base and Abstraction}
In enriched category theory, there is a notion of basis change, which maps $\mathcal{U}$-enriched categories to $\mathcal{V}$-enriched categories via a functor from $\mathcal{U}$ to $\mathcal{V}$.
In a similar fashion, computer scientists often employ a technique called \emph{abstract interpretation} which maps complicated concrete domains to simpler abstract domains modeled by lattices~\cite{CousotCousot1977}.
We conjecture that a carefully selected change of base could yield an abstract interpretation for parameterized circuits.

\bibliographystyle{eptcs}
\bibliography{generic}

\newpage
\appendix
\section{Basic Category Theory}
\label{Append:Cat}

A reoccurring theme in mathematics is to study structure preserving transformations between categories of mathematical objects.
For example, extension program semantics is the study of structure preserving transformations between a category of syntactic constructions, and some category of semantic interpretations.
A defining characteristic of category theory is that homomorphisms are seen as the defining characteristic of a category, rather than their specific objects.
Following this perspective, the key object of study in category theory is that of the functor, which describes homomorphisms between categories.

\begin{definition}[{\cite{MacLane2010}}]
    A \emph{(locally small) category} $\mathcal{C}$ consists of the following data.
    \begin{enumerate}
    \item \textbf{Objects}.
          A collection $\mathcal{C}_0$.
    \item \textbf{Morphisms}.
          For each $X, Y \in \mathcal{C}_0$, a set $\mathcal{C}( X, Y )$ whose elements are denoted $X \xrightarrow{f} Y$.
    \item \textbf{Identities}.
          For each $X \in \mathcal{C}_0$, a morphism $1_X \in \mathcal{C}( X, X )$.
    \item \textbf{Composition}.
          For each $X, Y, Z \in \mathcal{C}_0$, a function $\circ: \mathcal{C}( Y, Z ) \times \mathcal{C}( X, Y ) \to \mathcal{C}( X, Z )$.
    \end{enumerate}
    This data satisfies the following conditions.
    \begin{enumerate}
    \item \textbf{Associativity}.
          For each $X \xrightarrow{f} Y \xrightarrow{g} Z \xrightarrow{h} Z$, $h \circ ( g \circ f ) = ( h \circ g ) \circ f$.
    \item \textbf{Unitality}.
          For each $X \xrightarrow{f} Y$, $1_Y \circ f = f = f \circ 1_X$.
    \end{enumerate}
\end{definition}

\begin{example}
    Common categories include \textbf{Set} (sets and functions), \textbf{FVect} (finite-dimensional $\mathbb{C}$-vector spaces and linear transformations), \textbf{Top} (topological spaces and continuous functions), \textbf{Cat} (categories and functors) and \textbf{Pos} (posets and monotone maps).
\end{example}

\begin{example}
    If $( P, \le )$ is a poset, then there exists a category $\mathcal{P}$ such that $\mathcal{P}_0 = P$ and there exists a unique arrow from $x \in P$ to $y \in P$ if and only if $x \le y$.
\end{example}

A common trend in category theory is to define mathematical constructions by their data, and the equations it satisfies.
Often, these equations can become tedious to write out.
Instead, commuting diagrams are used to communicate the equations in a visual fashion.
A \emph{diagram} for a category $\mathcal{C}$ is a directed graph whose vertices are labeled by objects in $\mathcal{C}_0$, and whose edges are labeled by morphisms in $\mathcal{C}( X, Y )$ where $X$ is the label for the source of the edge and $Y$ is the label for the target of the edge.
A diagram is said to \emph{commute} if for each pair of paths with the same source and the same target, their labels compose to the same morphism (this is unambiguous by associativity).
For example, $X \xrightarrow{f} Y$ and $Y \xrightarrow{f} X$ are inverses if the following diagram commutes.
\begin{center}
\begin{tikzcd}
	& Y \\
	Y & X & X
	\arrow["{1_X}"', from=1-2, to=2-1]
	\arrow["g"', from=1-2, to=2-2]
	\arrow["f", from=2-2, to=2-1]
	\arrow["f"', from=2-3, to=1-2]
	\arrow["{1_X}", from=2-3, to=2-2]
\end{tikzcd}
\end{center}
Equationally, this says that $h \circ f = 1_X$ and $f \circ h = 1_Y$.
Note that when $h \circ f = 1_X$ we say that $f$ is a \emph{section} and $h$ is a \emph{retraction}.

\begin{definition}[\cite{MacLane2010}]
    A \emph{functor} $F: \mathcal{C} \to \mathcal{D}$ between categories is a choice of objects $F_0( X ) \in \mathcal{D}_0$ for each $X \in \mathcal{C}_0$, and an indexed family of functions $F_{X,Y}: \mathcal{C}( X, Y ) \to \mathcal{D}( F_0( X ), F_0( Y ) )$, satisfying the following conditions.
    \begin{enumerate}
    \item \textbf{Composition Preservation}.
          For each $X \xrightarrow{f} Y \xrightarrow{g} Z$, $F_{X,Z}( g \circ f ) = F_{Y,Z}( g ) \circ F_{X,Y}( f )$.
    \item \textbf{Identity Preservation}.
          For each $X \in \mathcal{C}_0$, $F_{X,X}( 1_X ) = 1_{F_0(X)}$.
    \end{enumerate}
    If $F_{X,Y}$ is inject for each $X, Y \in \mathcal{C}$, then $F$ is \emph{faithful}.
\end{definition}

\begin{definition}[{\cite{MacLane2010}}]
    A natural transformation $\eta: F \Rightarrow G$ between functors is an indexed family of morphisms $\eta_X: F_0( X ) \to G_0( X )$ such that the following diagram commutes.
    \begin{center}
\begin{tikzcd}
	{F_0(X)} & {G_0(X)} \\
	{F_0(Y)} & {G_0(Y)}
	\arrow["{\eta_X}", from=1-1, to=1-2]
	\arrow["{F(f)}"', from=1-1, to=2-1]
	\arrow["{F(g)}", from=1-2, to=2-2]
	\arrow["{\eta_Y}"', from=2-1, to=2-2]
\end{tikzcd}
    \end{center}
\end{definition}

\begin{example}
    Every category $\mathcal{C}$ admits an identity functor $1_{\mathcal{C}}: \mathcal{C} \to \mathcal{C}$ which acts as the identity both on objects and on moprhisms.
    There exists a functor $\Delta: \textbf{Set} \to \textbf{Set}$ such that $\Delta_0( X ) = X \times X$ and $\Delta_{X,Y}( f ) = ( x \mapsto ( f(x), f(x) )$.
    This concept will be generalized in the next section.
\end{example}

An important construction in the category of sets is the Cartesian product.
In set theory, the Cartestian product is usually defined explicitly using set-builder notation, as in~\cite{Jech2003,Hammack2009}.
However, the Cartesian product is also characterized by its projection maps, together with a universal property.
This characterization generalizes to other categories, as follows.

\begin{definition}[{\cite{MacLane2010}}]
    A category $\mathcal{C}$ has \emph{(finite) products} if for each pair of objects $X_1$ and $X_2$ in $\mathcal{C}$, there exists an object $X_1 \times X_2$ in $\mathcal{C}$ with morphisms $X_1 \times X_2 \xrightarrow{\pi_1} X_1$ and $X_1 \times X_2 \xrightarrow{\pi_2} X_2$ such that for each pair of morphisms $C \xrightarrow{f} X_1$ and $C \xrightarrow{g} X_2$ in $\mathcal{C}$, there exists a unique morphism $C \xrightarrow{h} X_1 \times X_2$ such that the following diagram commutes.
    \begin{center}
\begin{tikzcd}
	& C \\
	{X_1} & {X_1 \times X_2} & {X_2}
	\arrow["f"', from=1-2, to=2-1]
	\arrow["h", dashed, from=1-2, to=2-2]
	\arrow["g", from=1-2, to=2-3]
	\arrow["{\pi_1}", from=2-2, to=2-1]
	\arrow["{\pi_2}"', from=2-2, to=2-3]
\end{tikzcd}
    \end{center}
\end{definition}

\begin{example}
    The canonical example of a categorical product is the Cartesian product in \text{Set}.
    The categories \textbf{Top} and \textbf{FVect} also have finite products (the projections $\pi_1$ and $\pi_2$ are continuous).
    In \textbf{Top} this is given by the Cartesian product of topological spaces with the product topology.
    In \textbf{FVect} this is given by the direct product of vector spaces (the projections $\pi_1$ and $\pi_2$ are linear).
    In \textbf{Pos} this is given by the Cartesian product of the underlying set and order relation (the projections are trivially monotone).
\end{example}

\begin{example}[{\cite{MacLane2010}}]
    The product \textbf{Cat} also has finite products.
    This means that for each pair of categories $\mathcal{C}$ and $\mathcal{D}$ in \textbf{Cat}, there exists a product category $\mathcal{C} \times \mathcal{D}$.
    The objects in $( \mathcal{C} \times \mathcal{D} )_0$ are tuples of the form $( X, Y )$, where $X \in \mathcal{C}_0$ and $Y \in \mathcal{D}_0$.
    The morphisms in $\mathcal{C}( ( X, Y ), ( X', Y' ) )$ are the elements of $\mathcal{C}( X, Y ) \times \mathcal{D}( X', Y' )$.
    Composition of morphisms is defined component-wise.
    It follows that the identity morphism for $( X, Y )$ is $( 1_X, 1_Y )$.
    Moreover, there exists a functor $c_{\mathcal{C},\mathcal{D}}: \mathcal{C} \times \mathcal{D} \to \mathcal{D} \times \mathcal{C}$ such that $F_0( X, Y ) = ( Y, X )$ and $F_{X,Y}( f, g ) = ( g, f )$.
\end{example}

\begin{example}[{\cite{BluteScott2004}}]
    \label{Ex:Heyting}
    A poset $( P, \le )$ has products if and only if it is a semilattice.
    That is, there exists a meet $\land: P \times P \to P$ and a top element $\top \in P$ such that: (1) $a \le \top$ for all $a \in P$; (2) if $c \le a$ and $c \le b$ then $c \le a \land b$; (3) $a \le b \land a$ and $b \le a \land b$ for all $a \in P$ and $b \in P$.
    Moreover, if there exists an implication $\implies: P \times P \to P$ such that $c \land c \le b$ if and only if $c \le (a \implies b)$, then the poset is a model of intuitionistic propositional logic (also called a Heyting semilattice).
\end{example}

\section{Products and Monoidal Categories}
\label{Append:MonCat}

The categorical product describes a composition for objects in a category, which extends to all morphisms in the category.
However, the categorical product comes with far more data, such as the projection maps $\pi_1$ and $\pi_2$.
More generally, one can consider a monoid-like structure on the objects which extends to all morphisms in the category.
This generalization can be motivated by many constructions in mathematics, such as the Kronecker tensor product of vector spaces, and the parallel composition of circuits.
Both examples can be described as monoidal categories.

\begin{definition}[\cite{MacLane2010}]
    A \emph{monoidal category} $( \mathcal{C}, \otimes, \mathbb{I}, \alpha, \lambda, \rho )$ is a category $\mathcal{C}$ with the following data.
    \begin{enumerate}
    \item \textbf{Monoidal Product}.
          A functor $\otimes: \mathcal{C} \times \mathcal{C} \to \mathcal{C}$.
    \item \textbf{Monoidal Unit}.
          An object $\mathbb{I} \in \mathcal{C}_0$.
    \item \textbf{Associator}.
          A natural isomorphism $\alpha: ( - ) \otimes ( ( - ) \otimes ( - ) ) \Rightarrow ( ( - ) \otimes ( - ) ) \otimes ( - )$.
    \item \textbf{Unitors}.
          Natural isomorphism $\lambda: \mathbb{I} \otimes ( - ) \Rightarrow ( - )$ and $\rho: ( - ) \otimes \mathbb{I} \Rightarrow ( - )$.
    \end{enumerate}
    This data is subject to the condition that the following diagrams commute for all $X, Y, Z, W \in \mathcal{C}_0$.
    \begin{center}
\begin{tikzcd}[column sep=0.1em]
	& {( X \otimes Y ) \otimes ( Z \otimes W )} \\
	{X \otimes ( Y \otimes ( Z \otimes W ) )} && {( ( X \otimes Y ) \otimes Z ) \otimes W} \\
	{X \otimes ( ( Y \otimes Z ) \otimes W )} && {( X \otimes ( Y \otimes Z ) ) \otimes W}
	\arrow["{\alpha_{( X \otimes Y, Z, W )}}", from=1-2, to=2-3]
	\arrow["{\alpha_{( X, Y, Z \otimes W)}}", from=2-1, to=1-2]
	\arrow["{1_X \otimes \alpha_{( Y, Z, W )}}"', from=2-1, to=3-1]
	\arrow["{\alpha_{( X, Y \otimes Z, W )}}"', from=3-1, to=3-3]
	\arrow["{\alpha_{(X,Y,Z)} \otimes 1_W}"', from=3-3, to=2-3]
\end{tikzcd}
\begin{tikzcd}
	& {X \otimes Y} \\
	{X \otimes ( I \otimes Y )} & {( X \otimes I ) \otimes Y}
	\arrow["{1_X \otimes \lambda_Y}", from=2-1, to=1-2]
	\arrow["{\alpha_{X,I,Y}}"', from=2-1, to=2-2]
	\arrow["{\rho_X \otimes 1_Y}"', from=2-2, to=1-2]
\end{tikzcd}
    \end{center}
    A \emph{braiding} on $( \mathcal{C}, \otimes, \mathbb{I}, \alpha, \lambda, \rho )$ is a natural isomorphism $\beta: ( - ) \otimes ( - ) \Rightarrow ( ( - ) \otimes ( - ) ) \circ c$ such that the following diagrams commute for all $X, Y, Z \in \mathcal{C}_0$ (where $c$ is the braiding on \textbf{Cat}).
    \begin{center}
\begin{tikzcd}
	{X \otimes ( Y \otimes Z )} & {( X \otimes Y ) \otimes Z} & {Z \otimes ( X \otimes Y )} \\
	{X \otimes ( Z \otimes Y )} & {( X \otimes Z ) \otimes Y} & {( Z \otimes X ) \otimes Y}
	\arrow["{\alpha_{(X, Y, Z)}}", from=1-1, to=1-2]
	\arrow["{1_X \otimes \beta_{(Y,Z)}}"', from=1-1, to=2-1]
	\arrow["{\beta_{(X \otimes Y, Z)}}", from=1-2, to=1-3]
	\arrow["{\alpha_{(Z,X,Y)}}", from=1-3, to=2-3]
	\arrow["{\alpha_{(X,Z,Y)}}"', from=2-1, to=2-2]
	\arrow["{\beta_{(X,Z)} \otimes 1_Y}"', from=2-2, to=2-3]
\end{tikzcd}
\begin{tikzcd}
	{( X \otimes Y ) \otimes Z} & {X \otimes ( Y \otimes Z )} & {( Y \otimes Z ) \otimes X} \\
	{( Y \otimes X ) \otimes Z} & {Y \otimes ( X \otimes Z )} & {Y \otimes ( Z \otimes X )}
	\arrow["{\alpha_{(X,Y,Z)}^{-1}}", from=1-1, to=1-2]
	\arrow["{\beta_{(X,Y)} \otimes 1_Z}"', from=1-1, to=2-1]
	\arrow["{\beta_{(X, Y \otimes Z)}}", from=1-2, to=1-3]
	\arrow["{\alpha_{(Y,Z,X)}^{-1}}", from=1-3, to=2-3]
	\arrow["{\alpha_{(Y, X, Z)}^{-1}}"', from=2-1, to=2-2]
	\arrow["{1_Y \otimes \beta_{(X,Z)}}"', from=2-2, to=2-3]
\end{tikzcd}
    \end{center}
    If $\beta_{(X,Y)}^{-1} = \beta_{(Y,X)}$ for all $X, Y \in \mathcal{C}_0$, then $( \mathcal{C}, \otimes, \mathbb{I}, \alpha, \lambda, \rho, \beta )$ is a \emph{symmetric monoidal category (smc)}.
\end{definition}

\begin{example}[\cite{MacLane2010}]
    Every category with finite products and a terminal object (see~\cref{Def:Term}) is a monoidal category.
    In this case, the monoidal product maps each object $( X, Y ) \to X \times Y$ and each morphism $( f, g )$ to $f \times g$.
    In general, defining a functor $F: \mathcal{C} \times \mathcal{C} \to \mathcal{C}$ is challenging.
    In suffices to check that: (1) $F$ is functorial in the first component, that is $F( -, 1_X )$ is a functor for each $X \in \mathcal{C}_0$; (2) $F$ is functorial in the second component; (3) $F$ satisfies the \emph{interchange condition}, that is $F( i_{Y}, g ) \circ F( f, i_{X'}) = F( f, g ) = F( f, i_{Y'} ) \circ F( i_X, g )$ for all $f: X \to Y$ and $g: X' \to Y'$.
    For example, the Kronecker tensor product of vector spaces satisfies these properties.
    The category $\textbf{FVect}$ is monoidal with respect to the Kronecker tensor product and the monoidal unit $\mathbb{C}$.
\end{example}

\begin{definition}[{\cite{MacLane2010}}]
    Let $\mathcal{M} = ( \mathcal{C}, \otimes, \mathbb{I}, \alpha, \lambda, \rho )$ and $\mathcal{N} = ( \mathcal{D}, \boxtimes, \mathbb{J}, a, \ell, r )$ be monoidal categories.
    A \emph{(strict) monoidal functor} from $\mathcal{M}$ to $\mathcal{N}$ is a functor $F: \mathcal{C} \to \mathcal{D}$, satisfying the following conditions.
    \begin{enumerate}
    \item \textbf{Preserves Units}.
          $\mathbb{J} = F( \mathbb{I} )$.
    \item \textbf{Preserves Products}.
          If $X \xrightarrow{f} Y$ and $X' \xrightarrow{g} Y'$, then $F( f \otimes g ) = F( f ) \boxtimes F( g )$.
    \item \textbf{Preserves Structure}.
          For each $X, Y, Z \in \mathcal{C}_0$, $F( \alpha_{(X,Y,Z)} ) = a_{(X,Y,Z)}$, $F( \lambda_X ) = \ell_X$, and $F( \rho_X ) = r_X$.
    \end{enumerate}
    Moreover, when $\mathcal{C}$ admits a braiding $\beta$ and $\mathcal{D}$ admits a braiding $b$, then $F$ is a \emph{(strict) braided monoidal functor} when $F( \beta_{(X,Y)} ) = b_{(F(X), F(Y))}$ for each $X, Y \in \mathcal{C}_0$.
\end{definition}

The equations satisfied by the natural transformations in a monoidal category are referred to as \emph{coherence conditions}.
These ensure the uniqueness of structural morphisms composed from $\alpha$, $\lambda$, $\rho$, $\beta$, and the identities~\cite{Kelly1964,MacLane1963}.
More specifically, if there exists a structural morphism from an object $X$ to an object $Y$, then that morphism is unique.
This coherence leads to a graphical language in two dimensions, such that the horizontal dimension corresponds to morphism composition, and the vertical dimension corresponds to monoidal composition (see~\cite{Selinger2010}).
In these language, morphisms are depicted as circuits (called \emph{string diagrams}), with wires corresponding to objects and morphisms corresponding to gates.
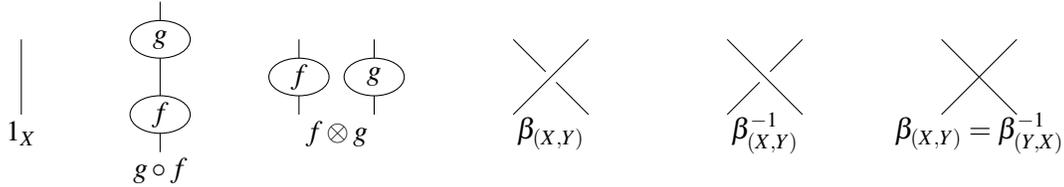
\begin{center}
\begin{tikzpicture}[baseline=(current bounding box.center),scale=0.5]
\useasboundingbox (-0.5,-0.5) rectangle (0.5,3.5);
\draw[,,] (0.00,1.00) -- (0.00,3.00);
\draw (0.00,0.50) node{$1_X$};
\end{tikzpicture}

    \qquad\quad
    \input{figs/hcomp}
    \qquad\quad
    \input{figs/vcomp}
    \qquad\quad
    \input{figs/braid}
    \qquad\quad
    \input{figs/invbraid}
    \qquad\quad
    \input{figs/sym}
\end{center}
The monoidal unit $\mathbb{I}$ is denoted by an invisible wire.
Each string diagram corresponds to an equivalence class of equations up to a choice of associators and unitors.
By labeling the boundaries of the string diagram (i.e., including the domains and codomains), all associators and unitors become fixed (up to coherence), and a unique morphism is selected~(see~\cite{JoyalStreet1991}).
These equivalence classes are preserved by appropriate isometries of the diagrams~(see~\cite{Selinger2010}).
In particular, monoidal isometries occur on the plane, braided monoidal isometries occur in three-dimensions (i.e., wires may pass over one-another), and symmetric monoidal isometries occur in four dimensions (i.e. wires may pass through one-another). 

The monoidal structure on \textbf{Set} defined by $\otimes = \times$ and $\mathbb{I} = \{ * \}$ has many nice properties.
For example, each set $X$ has a unique function $X \to \mathbb{I}$ defined by $x \mapsto *$, and each function $\mathbb{I} \to X$ picks out a unique element in $X$.
Moreover, for each object $X$, there exists a function $X \to X \times X$ defined by $a \mapsto ( a, a )$ which duplicates the elements in $X$.
These observations lead to two equivalent generalizations of this monoidal structure on \textbf{Set}, referred to as Cartesian categories.

\begin{definition}
    Let $( \mathcal{C}, \otimes, \mathbb{I}, \alpha, \lambda, \rho )$ be a monoidal category.
    A \emph{generalized element} of $X \in \mathcal{C}_0$ is a morphism $\mathbb{I} \to X$.
\end{definition}

\begin{definition}[\cite{MacLane2010}]
    \label{Def:Term}
    A \emph{terminal object} in $\mathcal{C}$ is an object $T \in \mathcal{C}_0$ such that for each $X \in \mathcal{C}_0$, there exists a unique morphism $X \xrightarrow{!} T$.
\end{definition}

\begin{definition}[{\cite{HeunenVicary2019}}]
    A \emph{Cartesian category} is a smc $( \mathcal{C}, \otimes, \mathbb{I}, \alpha, \lambda, \rho, \beta )$ such that $\otimes$ is the product functor for $\mathcal{C}$ and $\mathbb{I}$ is the terminal object for $\mathcal{C}$.
\end{definition}

\begin{definition}[\cite{HeunenVicary2019}]
    A smc $( \mathcal{C}, \otimes, \mathbb{I}, \alpha, \lambda, \rho, \beta )$ has \emph{uniform copying} if there exists a natural transformation $\Delta: ( - ) \Rightarrow ( - ) \otimes ( - )$ such that $\Delta_I = \rho_I$ and the following equations are satisfied for all $X, Y \in \mathcal{C}_0$.
    \begin{center}
        \input{figs/copy_1_lhs}
=
\input{figs/copy_1_rhs}

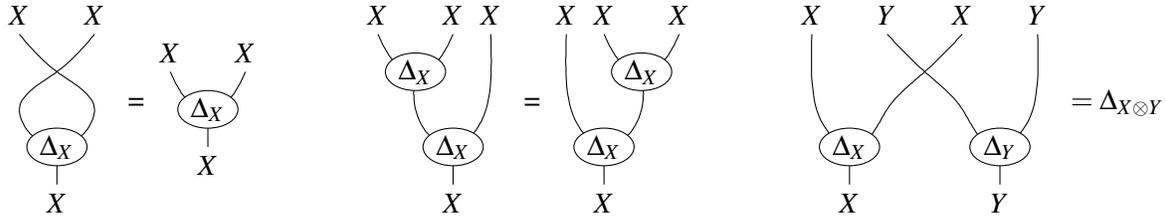
        \qquad\;
        \input{figs/copy_2_lhs}
=
\input{figs/copy_2_rhs}

        \qquad\;
        \input{figs/copy_3_lhs}
$= \Delta_{X \otimes Y}$
    \end{center}
\end{definition}

\begin{definition}[\cite{HeunenVicary2019}]
    A smc $( \mathcal{C}, \otimes, \mathbb{I}, \alpha, \lambda, \rho, \beta )$ has \emph{uniform deleting} if there exists a natural transformation $e: ( - ) \Rightarrow \mathbb{I}$ such that $e_\mathbb{I} = 1_\mathbb{I}$ and the following diagram commutes for all $X, Y \in \mathcal{C}_0$.
    
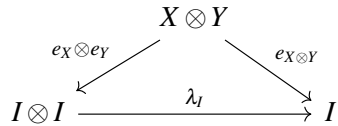
\begin{center}
\begin{tikzcd}
	& {X \otimes Y} \\
	{I \otimes I} && I
	\arrow["{e_X \otimes e_Y}"', from=1-2, to=2-1]
	\arrow["{e_{X \otimes Y}}", from=1-2, to=2-3]
	\arrow["{\lambda_I}", from=2-1, to=2-3]
\end{tikzcd}
    \end{center}
\end{definition}

\begin{proposition}[{\cite{HeunenVicary2019}}]
    Let $\mathcal{M} = ( \mathcal{C}, \otimes, \mathbb{I}, \alpha, \lambda, \rho, \beta )$ be a smc.
    Then $\mathcal{M}$ is a Cartesian category if and only if $\mathcal{M}$ has uniform copying and uniform deleting with the following equation satisfied for all $X \in \mathcal{C}_0$.
    \begin{center}
        \input{figs/cartesian_e1}
        =
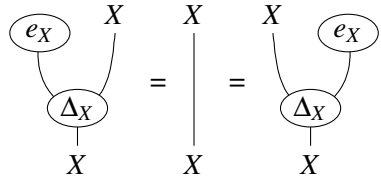
\begin{tikzpicture}[baseline=(current bounding box.center),scale=0.5]
\useasboundingbox (-0.5,-0.5) rectangle (0.5,5.5);
\draw[,,] (0.00,1.00) -- (0.00,4.00);
\draw (0.00,4.50) node{$X$};
\draw (0.00,0.50) node{$X$};
\end{tikzpicture}

        =
        \input{figs/cartesian_e2}
    \end{center}
\end{proposition}

It is important to note that the naturality of $\Delta$ states that $\Delta_Y \circ f = (f \otimes f) \circ \Delta_X$ for each $f \in \mathcal{C}( X, Y )$.
Moreover, if $Y = \mathbb{I}$, then $e_{I \otimes I} = \lambda_I$ and consequently $\Delta_Y \circ f = f \otimes f$.
It is in this sense that $\Delta$ captures the notion of uniform copying.
When uniform deleting is combined with a terminal unit, then it is also possible to ``\emph{delete}'' morphisms, since $e_Y \circ f = e_X$ and $e_\mathbb{I} = 1_{\mathbb{I}}$ by uniqueness.
These observations are captured by the following graphical equations.
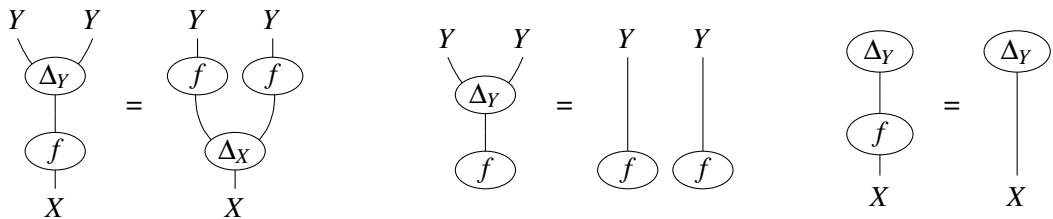
\begin{center}
    \input{figs/fun_copy_lhs}
=
\quad
\input{figs/fun_copy_rhs}

    \qquad\qquad
    \input{figs/fun_term_lhs}
=
\quad
\input{figs/fun_term_rhs}

    \qquad\qquad
    \input{figs/term_del_lhs}
\quad
=
\quad
\begin{tikzpicture}[baseline=(current bounding box.center),scale=0.55]
\useasboundingbox (-0.5,-0.5) rectangle (0.5,5.5);
\draw[,,] (0.00,4.00) -- (0.00,1.00);
\filldraw[fill=white] (0.00,4.00) ellipse (0.80cm and 0.50cm);
\draw (0.00,4.00) node{$\Delta_Y$};
\draw (0.00,0.50) node{$X$};
\end{tikzpicture}

\end{center}

\section{Diagrams for V-Monoidal Categories}
\label{Appendix:MonDiagrams}

Let $( \mathcal{C}, \otimes^{\mathcal{V}}, \alpha^{\mathcal{V}}, \lambda^{\mathcal{V}}, \rho^{\mathcal{V}} )$ be a $\mathcal{V}$-monoidal category.
First, let $X, Y, Z, X', Y', Z' \in \mathcal{C}^{\mathcal{V}}_0$.
Then the following equation captures composition preservation for $\otimes^{\mathcal{V}}$.
\begin{center}
    \input{figs/vprod_comp_lhs}
    \quad\;
    =
    \quad\;
    \input{figs/vprod_comp_rhs}
\end{center}
Next, let $X, Y \in \mathcal{C}^{\mathcal{V}}_0$.
Then the following equation captures identity preservation for $\otimes^{\mathcal{V}}$.
\begin{center}
    \input{figs/vprod_id_lhs}
    \quad\;
    =
    \qquad\qquad\;
    \begin{tikzpicture}[baseline=(current bounding box.center),scale=0.55]
\useasboundingbox (-0.5,-0.5) rectangle (0.5,4.5);
\draw[,,] (0.00,0.00) -- (0.00,3.00);
\filldraw[fill=white] (0.00,0.00) ellipse (0.80cm and 0.50cm);
\draw (0.00,3.50) node{$\mathcal{C}( X \otimes Y, X \otimes Y )$};
\draw (0.00,0.00) node{$1_{X \otimes Y}^{\mathcal{V}}$};
\end{tikzpicture}

\end{center}
Next, let $X, Y, Z, X', Y', Z' \in \mathcal{C}^{\mathcal{C}}_0$.
Write $A = ( X, Y, Z )$ and $B = ( X', Y', Z' )$.
Then the following equation captures the naturality of $\alpha^{\mathcal{V}}$.
\begin{center}
    \input{figs/vprod_assoc_lhs}
    \quad
    =
    \quad
    \input{figs/vprod_assoc_rhs}
\end{center}
Next, let $X, Y \in \mathcal{C}^{\mathcal{V}}_0$.
Then the following equation capture the naturality of $\lambda^{\mathcal{V}}$ and $\rho^{\mathcal{V}}$.
\begin{center}
    \input{figs/vprod_lunit_lhs}
    \quad\,
    =
    \quad\,
    \input{figs/vprod_lunit_rhs}
    \qquad
    \qquad
    \qquad
    \input{figs/vprod_runit_lhs}
    \quad\,
    =
    \quad\,
    \input{figs/vprod_runit_rhs}
\end{center}
Next, let $X, Y, Z, W \in \mathcal{C}^{\mathcal{V}}_0$.
Write $A = ( Y, Z, W )$, $B = ( X, Y \otimes Z, W )$, $C = ( X, Y ,Z )$, $D = ( X, Y, Z \otimes W )$, $E = ( X \otimes Y, Z, W )$, $S = X \otimes ( Y \otimes ( Z \otimes W ) )$, and $T = ( ( X \otimes Y ) \otimes Z ) \otimes W$.
Then the following equation captures the coherence condition for the associator.
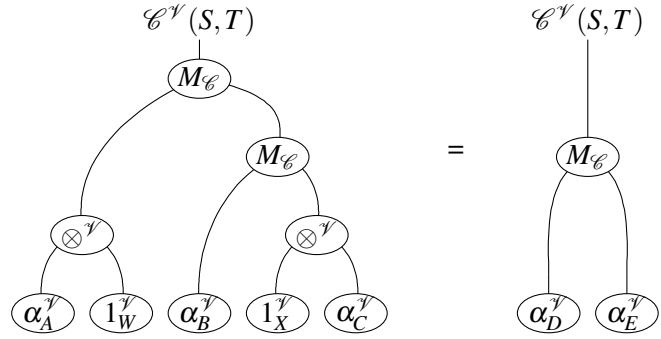
\begin{center}
    \input{figs/vprod_acohere_lhs}
    \qquad
    =
    \qquad
    \input{figs/vprod_acohere_rhs}
\end{center}
Finally, let $X, Y \in \mathcal{C}^{\mathcal{V}}_0$.
Write $A = ( X, \mathbb{I}, Y )$.
Then the following equation captures the coherence condition for the unitors.
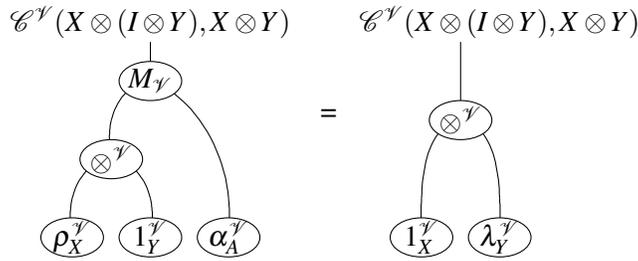
\begin{center}
    \input{figs/vprod_ucohere_lhs}
    \qquad  
    =
    \qquad
    \input{figs/vprod_ucohere_rhs}
\end{center}
Of course, $\alpha$, $\lambda$, and $\rho$ are not simply natural transformations, but rather natural isomorphisms.
This means that each component of $\alpha$, $\lambda$, and $\rho$ each has an inverse morphism.
These inverses behave the same as any $\mathcal{V}$-enriched inverse morphism.

\section{Diagrams for Braided V-Monoidal Categories}
\label{Appendix:BrDiagrams}

Let $( \mathcal{C}, \otimes^{\mathcal{V}}, \alpha^{\mathcal{V}}, \lambda^{\mathcal{V}}, \rho^{\mathcal{V}}, \beta^{\mathcal{V}} )$ be a braided $\mathcal{V}$-monoidal category.
First, let $X, Y, Z, X', Y', Z' \in \mathcal{C}_0^{\mathcal{V}}$.
Then the following equation captures composition preservation for $\otimes^{\textrm{rev}}$.
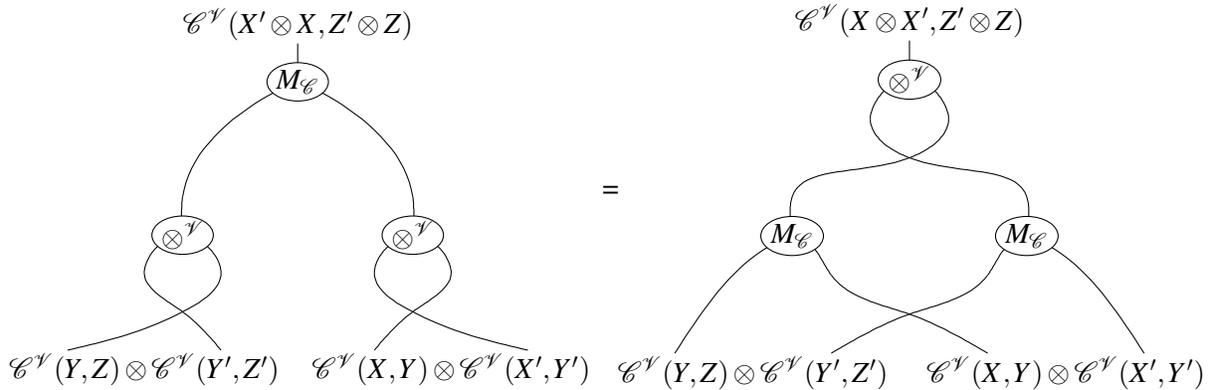
\begin{center}
    \input{figs/vprod_rev_comp_lhs}
    \quad\,
    =
    \quad
    \input{figs/vprod_rev_comp_rhs}
\end{center}
Next, let $X, Y \in \mathcal{C}_0^{\mathcal{V}}$.
Then the following equation captures identity preservation for $\otimes^{\textrm{rev}}$.
\begin{center}
    \input{figs/vprod_rev_id_lhs}
    \qquad
    =
    \qquad
    \begin{tikzpicture}[baseline=(current bounding box.center),scale=0.55]
\useasboundingbox (-0.5,-0.5) rectangle (0.5,6.5);
\draw[,,] (0.00,0.00) -- (0.00,5.00);
\filldraw[fill=white] (0.00,0.00) ellipse (0.80cm and 0.50cm);
\draw (0.00,5.50) node{$\mathcal{C}( Y \otimes X, Y \otimes X )$};
\draw (0.00,0.00) node{$1_{Y \otimes X}^{\mathcal{V}}$};
\end{tikzpicture}

\end{center}
Next, let $X, Y \in \mathcal{C}_0^{\mathcal{V}}$.
Then the following equation captures the naturality of $\beta^{\mathcal{V}}$ where $A = ( X, Y )$ and $B = ( X', Y' )$.
\begin{center}
    \input{figs/vbraid_lhs}
    \qquad
    =
    \qquad
    \input{figs/vbraid_rhs}
\end{center}
Next, let $X, Y, Z \in \mathcal{C}_0$.
Write $A = ( X, Z )$, $B = ( X, Z, Y )$, $C = ( Y, Z )$, $D = ( Z, X, Y )$, $E = ( X \otimes Y, Z )$, and $S = ( X, Y, Z )$.
Then the following equation captures the first coherence condition for $b$.
\begin{center}
    \input{figs/vbraid_cohere1_lhs}
    \qquad
    =
    \qquad\qquad
    \input{figs/vbraid_cohere1_rhs}
\end{center}
Finally, let $X, Y, Z \in \mathcal{C}_0$.
Write $A = ( X, Z )$, $B = ( Y, X, Z )$, $C = ( X, Y )$, $D = ( Y, Z, X )$, $E = ( X, Y \otimes Z )$, $S = ( X, Y, Z )$, and $\tau = \alpha^{-1}$.
Then the following equation captures the second coherence condition for $b$.
\begin{center}
    \input{figs/vbraid_cohere2_lhs}
    \qquad
    =
    \qquad\qquad
    \input{figs/vbraid_cohere2_rhs}
\end{center}
As with $\alpha$, $\lambda$, and $\rho$, $\beta$ is also a natural isomorphism.
This means that for each component of $\beta$, there exists an inverse $\mathcal{V}$-morphism with its corresponding diagrams.

\section{Proof of Lemma~\ref{Lem:BoxFunc}}

\begin{proof}
    First, it must be shown that $\boxtimes$ is functorial in the second component.
    Let $X \in \mathcal{C}_0$.
    Then for each morphism $f: Y \to Z$, the expression $i_X \boxtimes f$ simplifies as follows.
    \begin{equation}
        \label{Eq:1VarBox}
        i_X \boxtimes f
        =
        \quad\;\;\;
        \input{figs/box_simplify_1}
        \quad\;\;\;
        =
        \quad\;\;\;
        \input{figs/box_simplify_2}
        \quad\;\;\;
        =
        \quad\;\;\;
        \input{figs/box_simplify_3}
    \end{equation}
    To show composition preservation, let $X \xrightarrow{f} Y \xrightarrow{g} Z$.
    The following equation of diagrams holds.
    \begin{center}
        \input{figs/box_1var_comp_1}
        \;\;=\;\;
        \input{figs/box_1var_comp_2}
        \;\;=\;\;
        \input{figs/box_1var_comp_3}
        \;\;=\;\;
        \input{figs/box_1var_comp_4}
    \end{center}
    Then $( i_X \boxtimes g ) \star ( i_X \boxtimes f ) = i_X \boxtimes( g \star f )$.
    Since $f$ and $g$ were arbitrary, then $i_X \boxtimes ( - )$ preserves composition.
    To show identity preservation, let $Y \in \mathcal{C}_0$.
    Then the following equation of diagrams holds.
    \begin{center}
        $i_X \boxtimes i_Y$
        =
        \quad\;
        \input{figs/box_1var_id_1}
        \quad\;
        =
        \quad\;
        \input{figs/box_1var_id_2}
        \qquad
        =
        \qquad\;\;
        \input{figs/box_1var_id_3}
        \qquad\;\;
        =
        $i_{X \boxtimes Y}$
    \end{center}
    Since $Y$ was arbitrary, then $i_X \boxtimes ( - )$ preserves identities.
    Then $i_X \boxtimes ( - )$ is a functor.
    Since $X$ was arbitrary, then $\boxtimes$ is functorial in the second component.
    It follows by a symmetric argument that $\boxtimes$ is functorial in the first component.
    Finally, it must be shown that $\boxtimes$ satisfies the interchange condition.
    Let $f: X \to Y$ and $g: X' \to Y'$ in $\textbf{Param}( P, \mathcal{C} )$.
    Starting from $( i_{Y} \boxtimes g ) \star ( f \boxtimes i_{X'} )$, the composition in $\mathcal{C}^{\mathcal{V}}$ is pulled into $\otimes^{\mathcal{V}}$ and then simplified.
    \begin{center}
        \input{figs/box_functoriality_1}
        \quad\;
        =
        \quad\;
        \input{figs/box_functoriality_2}
        \quad\;
        =
        \quad\;
        \input{figs/box_functoriality_3}
    \end{center}
    Next, the two components of $\otimes$ are simplified to obtain $f \boxtimes g$.
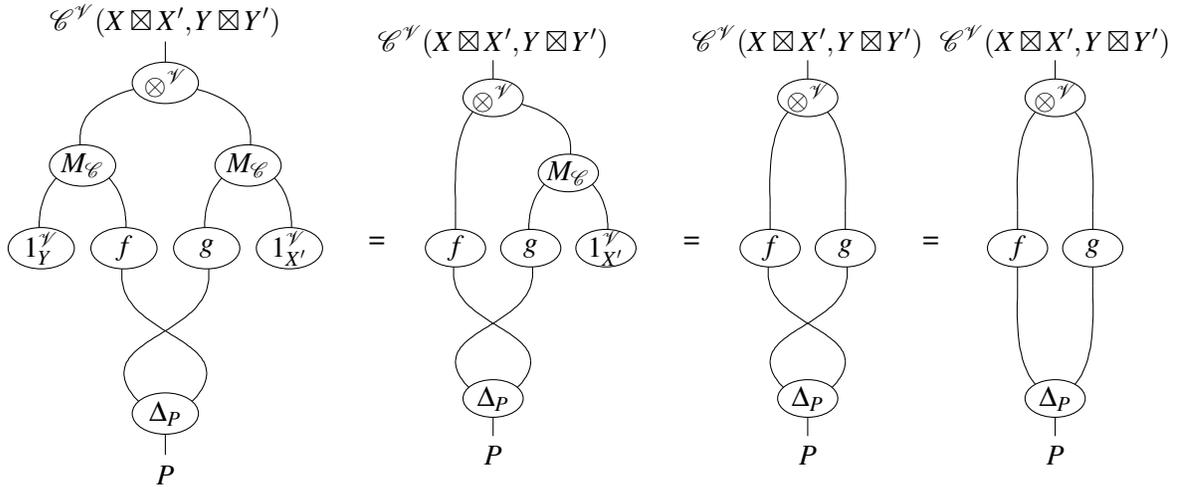
    \begin{center}
        \input{figs/box_functoriality_3}
        \quad\;
        =
        \quad\;
        \input{figs/box_functoriality_4}
        \quad\;
        =
        \quad\;
        \input{figs/box_functoriality_5}
        \quad\;
        =
        \quad\;\;
        \input{figs/box_functoriality_6a}
    \end{center}
    Then $( i_{Y} \boxtimes g ) \star ( f \boxtimes 1_{X'} ) = f \boxtimes g$.
    To obtain the second half of the interchange condition, identities are reintroduced on the opposite side of each component in $\otimes^{\mathcal{V}}$.
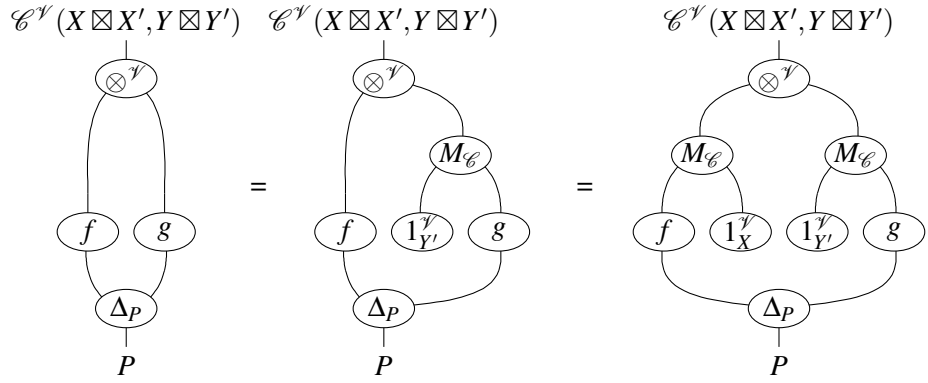
    \begin{center}
        \input{figs/box_functoriality_6b}
        \quad\;\;
        =
        \quad\;\;
        \input{figs/box_functoriality_7}
        \quad\;
        =
        \quad\;
        \input{figs/box_functoriality_8a}
    \end{center}
    Finally, $\otimes^{\mathcal{V}}$ can be distributed over the multiplication in $\mathcal{C}^{\mathcal{V}}$ to obtain $( f \boxtimes i_{Y'} ) \star ( i_{X} \boxtimes g )$.
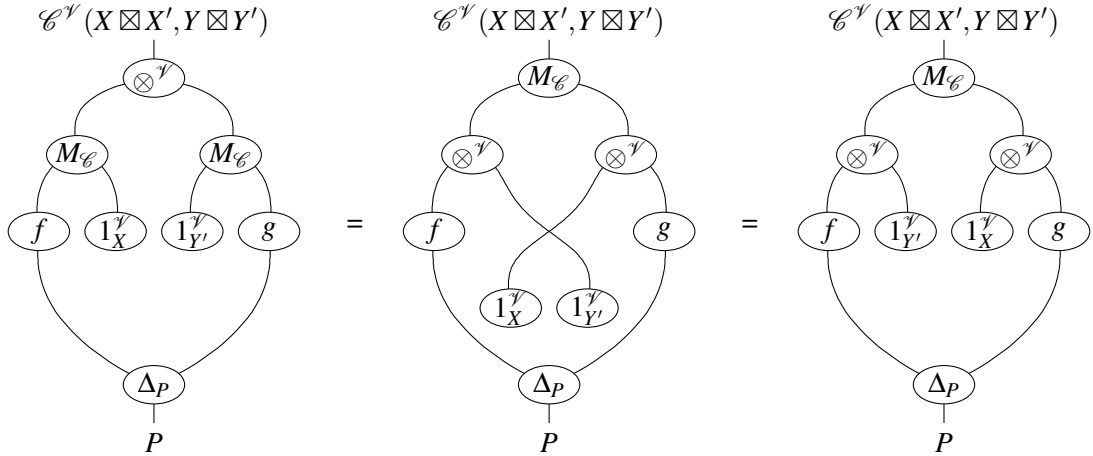
    \begin{center}
        \input{figs/box_functoriality_8b}
        \quad\;
        =
        \quad\;
        \input{figs/box_functoriality_9}
        \quad\;
        =
        \quad\;
        \input{figs/box_functoriality_10}
    \end{center}
    Since $f$ and $g$ were arbitrary, then $\boxtimes$ satisfies the interchange condition.
    In conclusion, $\boxtimes$ is a functor.
\end{proof}

\section{Proof of Lemma~\ref{Lem:BoxAssoc}}

\begin{proof}
    Let $X \xrightarrow{f} X'$, $Y \xrightarrow{g} Y'$, and $Z \xrightarrow{h} Z'$ in $\textbf{Param}( P, \mathcal{C} )$.
    Write $A = (X', Y', Z' )$ and $B =( X, Y, Z )$.
    Starting from $a_{A} \star ( f \boxtimes ( g \boxtimes h) )$, the term $a_A$ is expanded and simplified.
    \begin{center}
        \input{figs/box_assoc_1}
        \quad\;
        =
        \quad\;
        \input{figs/box_assoc_2}
        \quad\;
        =
        \quad\;
        \input{figs/box_assoc_3a}
    \end{center}
    Next, the associativity of $\otimes^{\mathcal{V}}$ is invoked via the naturality of $\alpha^{\mathcal{V}}$.
    \begin{center}
        \input{figs/box_assoc_3b}
        \quad\;
        =
        \quad\;
        \input{figs/box_assoc_4}
        \quad\;
        =
        \quad\;
        \input{figs/box_assoc_5a}
    \end{center}
    Finally, the $e_P$ term is reintroduced to recover $a_B$ on the right-hand side of the composite.
    \begin{center}
        \input{figs/box_assoc_5b}
        \quad\;
        =
        \quad\;
        \input{figs/box_assoc_6}
        \quad\;
        =
        \quad\;
        \input{figs/box_assoc_7}
    \end{center}
    Then $a_{A} \star ( f \boxtimes ( g \boxtimes h) ) = ( ( f \boxtimes g ) \boxtimes h ) \star a_{B}$.
    Since $f$, $g$ and $h$ were arbitrary, then $a$ is natural.
    It remains to be shown that $a$ is a natural isomorphism.
    Let $X, Y, Z \in \mathcal{C}_0$.
    Since $a_{(X,Y,Z)} = \alpha_{(X,Y,Z)} \circ e_P$ with $\alpha_{(X,Y,Z)}$ invertible, then $a_{(X,Y,Z)}$ is invertible by \cref{Lem:IncInv}.
    Since $X$, $Y$, and $Z$ were arbitrary, then $a$ is a natural isomorphism.
\end{proof}

\section{Proof of Lemma~\ref{Lem:BoxUnitors}}

\begin{proof}
    Let $f \in \textbf{Param}( P, \mathcal{C} )$.
    Starting from $\ell_Y \star ( i_{\mathbb{J}} \boxtimes f )$ and equation \cref{Eq:1VarBox}, the expression simplifies as follows.
    \begin{center}
        \input{figs/box_unitor_1}
        \quad
        =
        \quad
        \input{figs/box_unitor_2}
        \quad
        =
        \quad
        \input{figs/box_unitor_3}
    \end{center}
    Then by the naturality of $\lambda$, the monoidal product in the equation is eliminated as follows.
    \begin{center}
        \input{figs/box_unitor_3}
        \quad
        =
        \quad
        \input{figs/box_unitor_4}
        \quad
        =
        \quad
        \input{figs/box_unitor_5}
        \quad
        =
        \quad
        \input{figs/box_unitor_6}
    \end{center}
    Then $\ell_Y \star ( i_{\mathbb{J}} \boxtimes f ) = f \star \ell_{X}$.
    Since $f$was arbitrary, then $\ell$ is natural.
    It remains to be shown that $\ell$ is a natural isomorphism.
    Let $X \in \mathcal{C}_0$.
    Since $\ell_X = \lambda_X \circ e_P$ with $\lambda_X$ invertible, then $\ell_X$ is invertible by \cref{Lem:IncInv}.
    Since $X$ was arbitrary, then $a$ is a natural isomorphism.
    Then proof that $r$ is also a natural isomorphism is symmetric.
\end{proof}

\section{Proof of~Theorem~\ref{Thm:ParamMon}}

\begin{proof}
    By \cref{Lem:BoxFunc}, the data of $\boxtimes: \textbf{Param}( P, \mathcal{C} ) \times \textbf{Param}( P, \mathcal{C} ) \to \textbf{Param}( P, \mathcal{C} )$ defines a functor.
    Then by \cref{Lem:BoxAssoc} and \cref{Lem:BoxUnitors}, the bifunctor admits an associator $a$, a unit $\textbf{I}$, a left unitor $\ell$, and a right unitor $r$.
    It remains to be shown that $a$, $\ell$, and $r$ obey the coherence conditions of a monoidal category.
    First, the coherence for the associator is established.
    Let $X, Y, Z, W \in \mathcal{C}_0^{\mathcal{V}}$.
    Write $A = ( Y, Z, W )$, $B = ( X, Y \boxtimes Z, W)$, $C = ( X, Y, Z )$, $D = ( X, Y, Z \boxtimes W )$, $E = ( X \boxtimes Y, Z, W )$, $S = X \boxtimes ( Y \boxtimes ( Z \boxtimes W ) )$, and $T = ( ( X \boxtimes Z ) \boxtimes Z ) \boxtimes W$.
    Starting from $( a_A \star i_W ) \star ( a_B \star ( i_X \boxtimes a_C ) )$ and equation \cref{Eq:1VarBox}, the expression simplifies as follows.
    \begin{center}
        \input{figs/box_acohere_1}
        =
        \;
        \input{figs/box_acohere_2}
        $=\;\cdots\;=$
        \;
        \input{figs/box_acohere_3}
    \end{center}
    The top half of the diagram is then rewritten using the associator coherence from $\mathcal{C}^{\mathcal{V}}$.
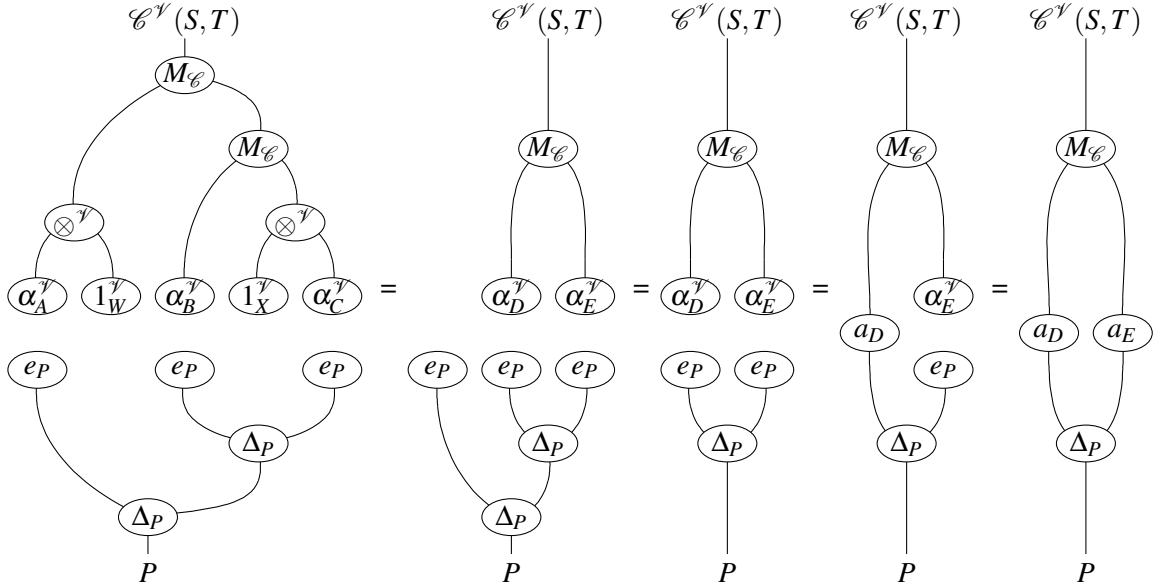
    \begin{center}
        \input{figs/box_acohere_3}
        \;
        =
        \;
        \input{figs/box_acohere_4}
        \;
        =
        \;
        \input{figs/box_acohere_5}
        \;
        =
        \;
        \input{figs/box_acohere_6}
        \;
        =
        \;
        \input{figs/box_acohere_7}
    \end{center}
    Then $( a_A \star i_W ) \star ( a_B \star ( i_X \boxtimes a_C ) ) = a_D \star a_E$.
    Since $A$, $B$, $C$, $D$, and $E$ were arbitrary, then coherence for the associator is established.
    It remains to establish coherence for the unitor.    
    Let $X, Y \in \mathcal{C}_0^{\mathcal{V}}$.
    Write $A = ( X, \mathbb{J}, Y )$, $S = X \boxtimes ( \mathbb{J}  \boxtimes Y )$, and $T = X \boxtimes Y$.
    Starting from $( r_X \boxtimes i_{\mathbb{J}} ) \star \alpha_A$ and equation \cref{Eq:1VarBox}, the expression simplifies as follows.
    \begin{center}
        \input{figs/box_ucohere_1}
        =
        \;
        \input{figs/box_ucohere_2}
        =
        \;
        \input{figs/box_ucohere_3}
        \;
        =
        \;
        \input{figs/box_ucohere_4}
        \;
        =
        \;
        \input{figs/box_ucohere_5}
        \;
        =
        \;
        \input{figs/box_ucohere_6}
    \end{center}
    Then $( r_X \boxtimes i_{\mathbb{J}} ) \star \alpha_A = i_{\mathbb{J}} \boxtimes \lambda_Y$.
    Since $X$ and $Y$ were arbitrary, then then coherence for the unitor is established.
    Since $a$, $\ell$, and $r$ satisfy the coherence conditions for a monoidal category, then $\boxtimes$, $\mathbb{J}$, $a$, $\ell$, and $r$ define a monoidal structure for $\textbf{Param}( P, \mathcal{C} )$ .
\end{proof}

\section{Proof of Lemma~\ref{Lem:BoxSym}}

\begin{proof}
    Let $f: X \to Y$ and $g: X' \to Y'$ in $\textbf{Param}( P, \mathcal{C} )$.
    Write $A = ( X, Y )$ and $B = ( X', Y' )$
    Starting from $( g \boxtimes f ) \star B_A$, the expression simplifies as follows.
    \begin{center}
        \input{figs/box_braid_1}
        \qquad
        =
        \qquad
        \input{figs/box_braid_2}
        \qquad
        =
        \qquad
        \input{figs/box_braid_3a}
    \end{center}
    Next, the naturality of $\beta$ is combined with the co-commutativity of $\Delta$ to transform the term $( - \boxtimes - )$ into the term $( - \boxtimes - ) \circ c$.
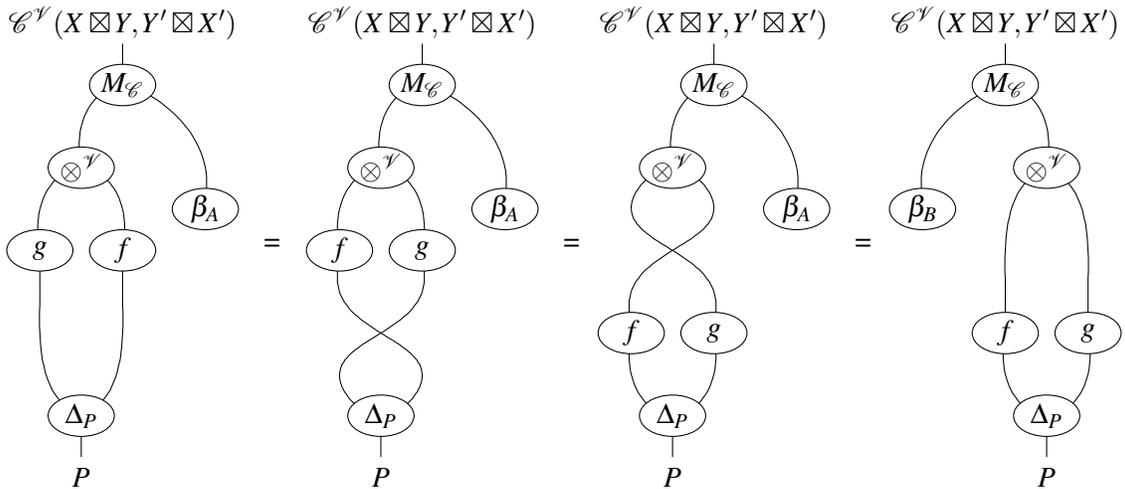
    \begin{center}
        \input{figs/box_braid_3b}
        \;\;
        =
        \;\;\;
        \input{figs/box_braid_4}
        \;\;
        =
        \;\;
        \input{figs/box_braid_5}
        \;\;
        =
        \;\;
        \input{figs/box_braid_6a}
    \end{center}
    Finally, the $b$-term is reintroduced to obtain $b_B \star ( g \boxtimes f )$.
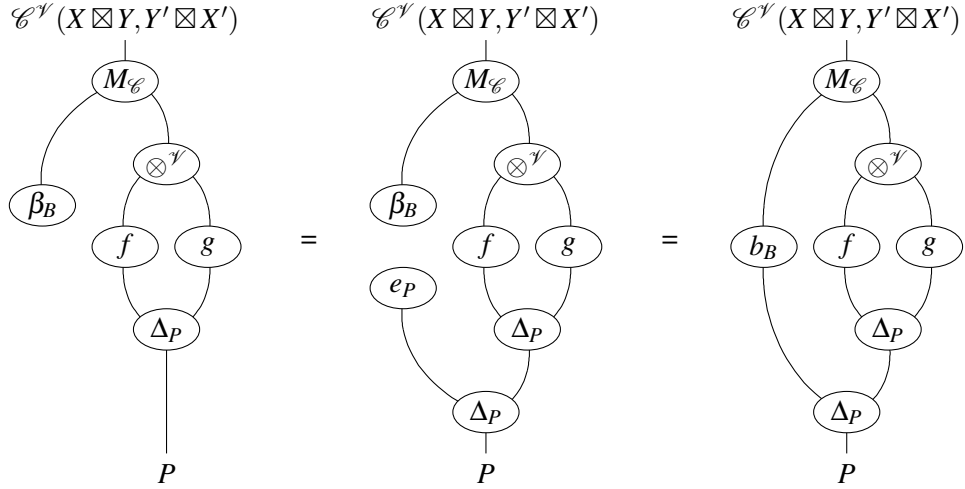
    \begin{center}
        \input{figs/box_braid_6b}
        \qquad
        =
        \qquad
        \input{figs/box_braid_7}
        \qquad
        =
        \qquad
        \input{figs/box_braid_8}
    \end{center}
    Since $f$ and $g$ were arbitrary, then $b$ is natural.
    Next, let $X, Y \in \mathcal{C}_0$.
    Since $b_{(X,Y)} = \beta_{(X,Y)} \circ e_P$ with $beta_{(X,Y)}$ invertible, then $\ell_X$ is invertible by \cref{Lem:IncInv}.
    Since $X$ and $Y$ were arbitrary, then $b$ is a natural isomorphism.
\end{proof}

\section{Proof of~Theorem~\ref{Thm:ParamBraid}}

\begin{proof}
    By \cref{Thm:ParamMon}, the data $( \textbf{Param}( P, \mathcal{C} ), \boxtimes, \mathbb{J}, a, \ell, r )$ defines a monoidal category.
    By \cref{Lem:BoxSym}, the data $b: ( - ) \boxtimes ( - ) \Rightarrow ( - ) \boxtimes ( - ) \circ c$ defines a natural isomorphism, where $c$ is the braiding for \textbf{Cat} as a Cartesian monoidal category.
    Then it suffices to show that $b$ satisfies the coherence conditions of a braided monoidal category.
    Let $X, Y, Z \in \mathcal{C}_0$.
    Write $A = ( X, Z )$, $B = ( X, Z, Y )$, $C = ( Y, Z )$, $D = ( Z, X, Y )$, $E = ( X \boxtimes Y, Z )$, $P = ( X, Y, Z )$, $S = X \boxtimes ( Y \boxtimes Z )$, and $T = ( Z \boxtimes X ) \boxtimes Y$.
    Starting from the expression $( b_A \star i_Y ) \star ( a_B \star ( i_X \boxtimes b_C ) )$ and equation \cref{Eq:1VarBox}, the terms simplify as follows.
    \begin{center}
        \input{figs/box_bcohere_1}
        =
        \;
        \input{figs/box_bcohere_2}
        $=\;\cdots\;=$
        \;
        \input{figs/box_bcohere_3}
    \end{center}
    The top half of the diagram is then rewritten using the first braiding coherence from $\mathcal{C}^{\mathcal{V}}$ to obtain the expression $\alpha_D \star ( \beta_E \star \alpha_P )$.
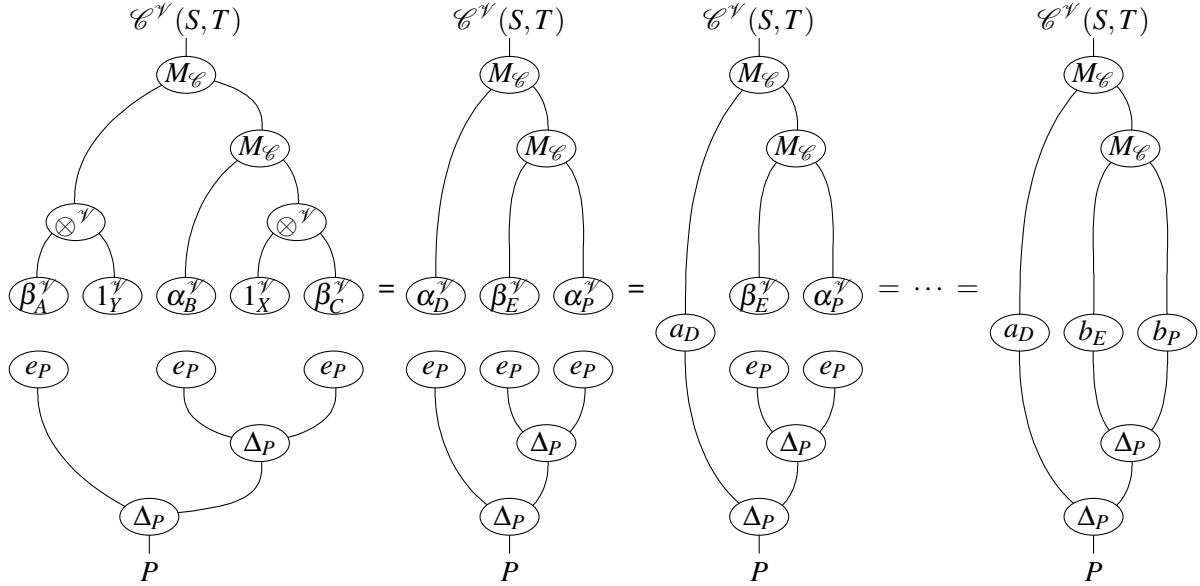
    \begin{center}
        \input{figs/box_bcohere_3}
        \,
        =
        \;
        \input{figs/box_bcohere_4}
        \,
        =
        \;
        \input{figs/box_bcohere_5}
        \,
        $=\;\cdots\;=$
        \;
        \input{figs/box_bcohere_6}
    \end{center}
    Then $( b_A \star i_Y ) \star ( a_B \star ( i_X \boxtimes b_C ) ) = \alpha_D \star ( \beta_E \star \alpha_P )$.
    Since $X$, $Y$, and $Z$ were arbitrary, then $b$ satisfies the first coherence condition of a braided monoidal category.
    The proof that $b$ satisfies the second coherence condition follows in a similar fashion.
    In conclusion, $( \textbf{Param}( P, \mathcal{C} ), \boxtimes, \mathbb{J}, a, \ell, r, b )$ is a braided monoidal category.
\end{proof}

\section{Proof of~Theorem~\ref{Thm:SemIncl}}

\begin{proof}
    First, it will be shown that $j$ is a functor.
    Let $X \xrightarrow{f} Y \xrightarrow{g} Z$ in $\mathcal{C}$.
    Then the following equation of diagrams hold.
    \begin{center}
        $j( g \circ f )$
        =
        \qquad
        \input{figs/j_fun_comp_lhs}
        \qquad
        =
        \qquad
        \input{figs/j_fun_comp_rhs}
        \qquad
        =
        $j( g ) \star j( f )$
    \end{center}
    Since $f$ and $g$ were arbitrary, then $j$ preserves composition.
    Next, let $X \in \mathcal{C}$.
    It follows by definition that $j( 1_X ) = j( 1_X^{\mathcal{V}} ) = 1_X^{\mathcal{V}} \circ e_P = i_X$.
    Since $X$ was arbitrary, then $j$ preserves identities.
    Then $j$ is a functor.

    Next, assume that $\mathcal{C}^{\mathcal{V}}$ admits $\mathcal{V}$-monoidal structure $( \mathcal{C}^{\mathcal{V}}, \otimes^{\mathcal{V}}, \mathbb{J}^{\mathcal{V}}, \alpha^{\mathcal{V}}, \lambda^{\mathcal{V}}, \rho^{\mathcal{V}} )$.
    It will be shown in this case that $j$ is a strict monoidal functor.
    Clearly $j$ preserves the monoidal unit, since $j_0$ is the identity.
    To show that $j$ preserves monoidal products, let $f: X \to Y$ and $g: X' \to Y'$ in $\mathcal{C}$.
    Then the following equation of diagrams hold.
    \begin{center}
        $j( g \otimes f )$
        =
        \qquad
        \input{figs/j_mon_comp_lhs}
        \qquad
        =
        \qquad
        \input{figs/j_mon_comp_rhs}
        \qquad
        =
        $j( g ) \boxtimes j( f )$
    \end{center}
    Since $f$ and $g$ were arbitrary, then $j$ preserves monoidal products.
    Next, let $X, Y, Z \in \mathcal{C}_0$.
    It follows by definition that $j( \alpha_{(X,Y,Z)} ) = a_{(X,Y,Z)}$, $j( \lambda_X ) = \ell_X$, and $j( \rho_X ) = r_X$.
    Since $X$, $Y$, and $Z$ were arbitrary, then $j$ preserves the structural natural transformations of $\otimes$.
    Then $j$ is a strict monoidal functor.
    
    Next, assume that $( \mathcal{C}^{\mathcal{V}}, \otimes^{\mathcal{V}}, \mathbb{J}^{\mathcal{V}}, \alpha^{\mathcal{V}}, \lambda^{\mathcal{V}}, \rho^{\mathcal{V}} )$ admits a $\mathcal{V}$-braiding $\beta^{\mathcal{V}}$.
    Let $X, Y \in \mathcal{C}_0$.
    Then $j( \beta_{(X,Y)} ) = b_{(X,Y)}$ by definition.
    Since $X$ and $Y$ were arbitrary, then $j$ preserves the braiding.
    Therefore, $j$ is a strict braided monoidal functor.

    Finally, assume that $\mathcal{V}( \mathbb{I}, P ) \ne \varnothing$.
    Then there exists a $\theta \in \mathcal{V}( \mathbb{I}, P )$.
    Let $X, Y \in \mathcal{C}_0$.
    Let $f, g: X \to Y$.
    Assume that $j( f ) = j( g )$.
    Then $\textsf{ev}_\theta( j( f ) ) = \textsf{ev}_\theta( j( g ) )$.
    Moreover, $\textsf{ev}_\theta( j( f ) ) = f$ and $\textsf{ev}_\theta( j( g ) ) =g$ by \cref{Lem:InclEv}.
    Then $f = g$.
    Since $f$ and $g$ were arbitrary, then $j_{X,Y}$ is injective.
    Since $X$ and $Y$ were arbitrary, then $j$ is faithful.
\end{proof}

\end{document}